\documentclass[letterpaper,11pt]{article}
\pdfoutput=1 

\usepackage{jcappub} 
\usepackage[T1]{fontenc} 

\newcommand{\sigmapsi}{$\sigma_{\mathrm{p}}^{\mathrm{SI}}$}
\newcommand{\sigmapsie}{\sigma_{\mathrm{p}}^{\mathrm{SI}}}

\newcommand{\Mvir}{M$_{\textrm{vir}}$}

\title{\boldmath The highest-speed local dark matter particles come from the Large Magellanic Cloud}

\author[a,1]{G. Besla,\note{Corresponding Author}}
\author[b]{A. H. G. Peter,}
\author[a]{N. Garavito-Camargo}

\affiliation[a]{Steward Observatory, University of Arizona, 933 North Cherry Avenue, Tucson, AZ 85721, USA}
\affiliation[b]{CCAPP, Department of Physics, and Department of Astronomy, The Ohio State University, 191 W. Woodruff Ave., Columbus, OH 43210, USA}

\emailAdd{gbesla@email.arizona.edu}
\emailAdd{peter.33@osu.edu}
\emailAdd{jngaravitoc@email.arizona.edu}

\abstract{Using N-body simulations of the Large Magellanic Cloud (LMC's) passage through the Milky Way (MW), tailored to reproduce observed kinematic properties of both galaxies, we show that the high-speed tail of the Solar Neighborhood dark matter distribution is 
overwhelmingly of LMC origin. Two populations contribute at high speeds: 1) Particles that were once bound to the LMC, and 2) MW halo particles that have been accelerated owing to the response of the halo to the 
recent passage of the LMC.
These particles reach speeds of 700-900 km/s with respect to the Earth, above the local escape speed of the MW.
The high-speed particles follow
trajectories similar to the Solar reflex motion, with peak velocities reached in June.
For low-mass dark matter, these high-speed particles can dominate the signal in direct-detection experiments, extending the reach of the experiments to lower mass and elastic scattering cross sections even with existing data sets.  Our study shows that even non-disrupted MW satellite galaxies can leave a significant dark matter footprint in the Solar Neighborhood. 

}

\begin{document}
\maketitle
\flushbottom

\section{Introduction}
\label{sec:intro}

The nature of dark matter (DM) is one of the fundamental unsolved mysteries in astrophysics. 
The simplest phenomenological model of DM, the cold dark matter (CDM) paradigm of a cold, collisionless, stable particle species, works remarkably well to describe the expansion of the Universe and the web of cosmic structure \citep{Ade:2015,Abbott:2017,Troxel:2017}.  A class of particle CDM candidate, the weakly interacting massive particle (WIMP; \cite{Steigman:1984}), naturally reproduces DM at the abundance measured precisely with cosmological measurements, on account of a small coupling between DM and baryonic matter \citep{Ade:2015}.  This small coupling motivates direct and indirect astroparticle searches for WIMPs in the Milky Way (MW) and beyond.  And yet, WIMP CDM has never been detected in an experiment.  A confirmation of the WIMP CDM paradigm requires observing a consistent signal in astroparticle experiments and astronomical observations \citep{Buckley:2017ijx}, but an assessment of consistency requires accurate theoretical modeling of our Galaxy's DM halo and any massive perturbers. 

A prime example of the intertwining of signals in astroparticle experiments and the distribution of DM in the MW is the interpretation of direct-detection experiments.  If DM consists of WIMPs (or axions \cite{Peccei:1977}), laboratory experiments directly detect DM as it passes through the apparatus \cite{Sikivie:1983ip,Goodman:1984dc,Drukier:1986tm,Wasserman:1986hh,VanBibber:1987rq}.  The number of events, the energy spectrum of recoiling nuclei/electrons (WIMPs) or photons (axions), and their angular distribution depend on the Solar Neighborhood DM phase-space density as well as DM's intrinsic particle properties.  While there are well-motivated ideas about what the Solar Neighborhood DM phase-space density might be (e.g., Refs.  \cite{Pato:2015dua,Necib:2018igl,Evans:2019bqy}), the only way to understand and measure it will be via these same direct-detection experiments \citep{Peter:2013,OHare:2017}.

This presents both a problem and opportunity. The problem is that, in the absence of a secure detection, the experiments require an assumption about the DM density and its velocity distribution in the Solar Neighborhood in order to interpret any potential signal in terms of DM particle-physics parameters (e.g., mass, cross sections; for a novel alternative ``halo-free'' method, see Refs. \cite{Blennow:2015oea,Ferrer:2015bta}).  Current input models, however, are unrealistically simple, frequently invoking a Maxwell-Boltzmann velocity distribution (usually assumed to be isotropic, but sometimes anisotropic \cite{Green:2002a}), and an axisymmetric or triaxial mass distribution of DM that is in dynamical equilibrium \citep{Peter:2009mi,Catena2012,Strigari:2013iaa,Read:2014qva,Pato:2015dua}.  The isotropic Maxwell-Boltzmann model with a Solar Neighborhood DM mass density of $\rho_\chi = (0.3-0.4)$ GeV/cm$^3$ and a cut-off at the escape speed from the MW is called the ``standard halo model'' (SHM) \cite{Lewin:1995rx,Green:2017odb}.  A number of simulations, though, indicate that the SHM is a poor model for the MW, even if it allows for convenient benchmarking of experiments against each other at the present (e.g., Refs. \cite{Lisanti:2011as,Sloane:2016,Bozorgnia:2016,Mao:2014}. 
The inclusion of baryons may bring the local DM speed distribution closer to Maxwellian \cite{Vogelsberger2009,Kuhlen:2010,Kelso:2016}, although with significant halo-to-halo variations.

The \textit{opportunity} is that the phase-space density, shaped by effects of the MW disk and the recent accretion history of the MW, may be directly measured from WIMP detections (e.g., Ref.~\cite{Peter:2013}).  Both effects are only partially theoretically and observationally understood. 
At geocentric speeds of $\sim 200$ km/s$-300$ km/s, the disk of the MW likely has a dramatic effect on the Solar Neighborhood DM density and velocity distribution (e.g., Refs.~\cite{Kuhlen:2013tra,Sloane:2016, deSalas:2019}), although other authors have argued the opposite (see, e.g., \cite{Kelso:2016,Bozorgnia:2016}).  A measurement of this effect would provide an important constraint on hydrodynamic simulations of the co-evolution of galactic disks and their DM halos.

The satellites of the MW---both the ones identified as galaxies as well as those dissolving into the stellar halo---are evidence of the MW's non-equilibrium state and assembly history, and should be imprinted in the Solar Neighborhood DM phase-space density. The cosmological assembly history of a galaxy can induce significant halo-to-halo scatter in the velocity distribution function \citep{Mao:2013,Mao:2014}. Several studies associate DM with un-phase-mixed stellar debris, notably: Sagittarius stellar stream \citep{LyndenBell:1995zz,Newberg:2001sx}, Gaia-Enceladus \citep{Helmi2018,Kruijssen2019}, Gaia-Sausage \citep{Belokurov2018}, Sequoia \& S1 stream \citep{Myeong2019, Myeong2018}, and Nyx \cite{Necib2019}. The DM associated with these stellar substructures is expected to 
leave distinct signatures in direct-detection experiments \cite{Freese:2003na,Freese:2003tt,Purcell:2012sh,Evans:2019bqy,OHare:2017,OHare:2018trr,Necib:2019iwb,Buckley:2019}.  It is possible, though unlikely, that the solar system is sitting inside a dense subhalo, which would significantly influence the local DM phase-space density \cite{Stiff2001,Helmi2002,Kamionkowski2008}. 

In current models, the highest-speed DM particles  
are spatially homogenized remnants of recently accreted and disrupted subhalos  \citep{Kuhlen:2010,Kuhlen:2012,Necib:2018igl}.
For low-mass WIMPs, these high-speed particles can {\it dominate} the signal, since only the fastest particles are energetic enough to reach the experimental threshold.

Current estimates of both effects -- the disk effects on the moderate-speed population and the streams, and the substructure origin of the high-speed tail -- come from a handful of hydrodynamic simulations, none of which 
contain an analog of the most massive substructure orbiting within the MW halo, the Large Magellanic Cloud (LMC).
An appropriate analog of the LMC must have 3D velocity and position 
vectors that are consistent with the observations of the real system, a feat that is not achievable in cosmological simulations. This omission is problematic, as the LMC is the most massive and most recent merger event for the MW, with a recent high-speed pericenter passage at a distance of $\sim $48 kpc only $\sim$50 Myr ago \cite{Besla:2007}. 

In this study, our models account for three effects simultaneously: 1) the MW disk potential; 2) the MW DM halo velocity anisotropy profile; and, for the first time, 3) the presence of the LMC.  At 50 kpc, the LMC is a significantly more massive, higher speed and closer satellite than is found in the majority of cosmological simulations \citep{Boylan:2011,Patel:2017}. 
In the following we illustrate that the high-speed tail of the local DM distribution is 
overwhelmingly of LMC origin -- both as particles that were once bound to the LMC and 
MW halo particles that have been accelerated owing to halo resonances induced by 
the motion of the LMC \cite{G19}.

\section{The LMC: A Massive Satellite on First Infall}\label{sec:massivesat}

The influence of the LMC on the DM structure of the MW has been ignored in all existing models of the MW's DM potential. However, two new pieces of information about the LMC are causing the astronomical community to reconsider its omission:

{\bf 1) The LMC is just past its first close approach to the MW:} Recent measurement of the 3D velocity of the LMC has revealed that it is moving too fast to be a long-term companion of our Galaxy \citep{Kallivayalil:2013}.  It is expected to have only recently ($<$ 1-2 Gyr) entered our neighborhood and is just past its only pericentric approach to the MW in at least 6 Gyr \citep{Besla:2007}. The LMC's closest approach to the MW occurred only
$\sim$50 Myr ago at a distance of $\sim$ 48 kpc.
This first infall scenario is cosmologically expected
for high-speed satellites of MW-like galaxies \citep{Boylan:2011,Busha:2011,Patel:2017}
and implies that the LMC should have retained a significant fraction of the DM mass it had upon entering the MW halo \citep{G19}.

{\bf 2) The LMC is a Massive Satellite:}
The stellar mass of the LMC is $\sim 2.7 \times 10^9$ M$_\odot$ \citep{van:09}.  
Abundance matching techniques indicate that the LMC's cosmologically expected halo mass at infall is  $\sim (1-3) \times 10^{11}$ M$_\odot$ (e.g., Ref.~\cite{Moster:2013}).  This is $10\times$ larger than conventionally assumed for the LMC. Recently, it was shown that such a high LMC mass will not perturb the MW's disk to levels inconsistent with observations \citep{Laporte:2018} and is both supported by the timing argument \citep{Penarrubia:2016} and needed to explain the kinematics of the Orphan Stream \citep{Erkal:2019} (and likely many other stellar streams whose kinematics are now measured by Gaia \citep{Shipp:2019}).
Furthermore, such high infall mass
prevents the LMC disk from being tidally truncated by the MW, explaining new results that the LMC's outer stellar disk extends to a radius approximately half its current separation
to the MW ($\sim 18.5$ kpc; \citep{Mackey:2016, Besla:2016}). 
Significantly, Ref. \cite{Shao:2018} finds that
satellites of MW type galaxies 
with stellar masses similar to the LMC and with a satellite companion
comparable in mass to the Small Magellanic Cloud (SMC), are found to have a pre-capture DM mass of
M$_{\textrm{vir}}=3.4^{+1.8}_{-1.2} \times 10^{11} $M$_{\odot}$. This high mass
is also 
supported by studies of cosmological dwarf galaxy pairs in the field \citep{Besla:2018}.

The mounting evidence supporting the LMC's high infall mass
implies that the LMC cannot be treated as a point mass tracer of the halo potential and cannot be ignored in dynamical models of the MW (see the review in Refs.~\cite{G19,Besla:2015}). Rather, such a massive satellite will contribute to the DM distribution of the MW. In particular, the virial radius of the LMC at infall is expected to be of order 100 kpc.  Given its current separation of 50 kpc, this implies that the LMC's DM halo overlaps with the MW's disk and our own Solar Neighborhood. In this study, we explore the contribution of DM particles
from the LMC's halo to the density and kinematics of DM 
in the Solar Neighborhood in a first infall scenario.


\section{Numerical Methods}
\label{sec:Numerical}
The LMC is just past its first pericentric approach to the MW \citep{Kallivayalil:2013, Besla:2007}. This is a rare orbital configuration and cosmological analogs are difficult to identify at z=0 \citep{Patel:2017}.
Moreover, because the kinematic properties of LMC particles in the Solar Neighborhood 
will depend on the exact orbital motion of the LMC through the halo, we cannot turn to cosmological simulations to study the LMC's impact on the velocity distribution of DM in
the Solar Neighborhood. Instead, we utilize idealized N-body simulations of the LMC-MW encounter presented in Ref. \cite{G19} (hereafter, G19), which we summarize briefly here.  

All simulations are created using the N-body/TreePM Smoothed Particle Hydrodynamics (SPH) code P-GADGET-3
an updated version of the publicly available GADGET-2 code \citep{Springel:2005}. N-body galaxy models are generated using the publicly available GALIC code, which implements an iterative method for the construction of N-body galaxy models in collisionless equilibrium \citep{Yurin:2014}.

\subsection{The MW Model}
 \label{sec:MW}
 
 The MW is modeled as a Hernquist DM halo \citep{Hernquist:1990} of 
 mass \Mvir $= 1.2 \times 10^{12}$ M$_\odot$. The virial
mass \Mvir~is defined as the mass enclosed within the virial radius, 
R$_{\textrm{vir}}$; the radius where the halo density 
$\rho_{\textrm{vir}} = \Delta_{\textrm{vir}} \Omega_m \rho$, with $\rho$ being
the average density of the Universe. The equivalent halo mass for the Hernquist
profile is M$_{\rm Hern} = 1.56\times 10^{12}$ M$_\odot$.
The circular speed for the MW DM halos in both 
Model 1 and Model 2 (see below) peaks at $V_{\rm max} = 208$ km/s. 

The MW's DM halo is simulated with $\sim 10^8$ particles, providing a DM mass resolution of $4 \times 10^4$ M$_\odot$. This is comparable to state-of-the-art cosmological simulations, like Caterpillar ($\sim 10^4$ M$_\odot$) \cite{Griffen2016}, ELVIS ($\sim 1.9 \times 10^5$ M$_\odot$) \cite{GarrisonKimmel2014} 
 LATTE ($\sim 3.5 \times 10^4$ M$_\odot$) \cite{Wetzel2016} or APOSTLE ($\sim 5\times 10^4$ M$_\odot$) 
 \cite{Sawala:2016}.
 By modeling the MW DM halo using as a spherical Hernquist distribution, any velocity
 asymmetries that result can be unambiguously attributed to the action of the LMC.

 The stellar component of the MW is modeled as a live 
 Hernquist bulge (M$_* = 10^{10}$ M$_\odot$) and exponential disk (M$_* = 5.78 \times 10^{10}$ M$_\odot$). Live stellar components allow us to appropriately account for 
 gravitational accelerations from MW baryons and also allows us to accurately track 
 the center of mass motion of the MW and the relative location of the Sun. 

 The properties of the MW disk, bulge and halo (e.g. scale lengths, mass) are 
 listed in G19 (their table 1) and are chosen such that the disk 
 kinematics roughly reproduce the observed
 rotation curve of the MW, which peaks at a value of $V \sim 240$ km/s \citep{McMillan:2011, McMillan:2017}.
 Note that this 
 circular speed is higher than generally assumed in the Standard Halo Model
 (e.g., Ref.~\cite{Herzog:2018}).

\subsection{Initial MW Halo Kinematics}
\label{sec:initialkinematics}

In order to fully specify the initial conditions for the DM halo particles in the simulations, we must additionally make assumptions about the velocity distribution.  The velocity distribution can either be isotropic --- the velocity dispersion is the same in every direction, like for an ideal gas --- or anisotropic.  While idealized simulations often adopt isotropy for simplicity (one may then use Eddington's formula to define the full 6D phase-space density given a spatial density profile \cite{Binney:2008}), real halos typically have significant radial anisotropy resulting from their accretion histories.  At a given radius in the halo, the so-called
\textit{anisotropy parameter}, $\beta$,  is defined as:

\begin{equation}
\beta = 1 - \dfrac{\sigma_t^2}{2\sigma_r^2} .
\end{equation}

Following G19, 
we explore two MW halo models with different initial velocity 
anisotropy profiles,
$\beta(r)$: 1) \textbf{Model 1} an isotropic 
DM halo ($\beta(r)=0$); and 2) \textbf{Model 2} a radially
biased profile~\citep{Hansen:2006}, consistent with the kinematics of DM
halos formed in cosmological
simulations~\cite{Abadi:2006}, 

\begin{equation}
\beta(r)=-0.5-0.2\alpha(r) ; 
\qquad
\alpha(r)=\dfrac{d \ln \rho_{DM}}{d \ln r}.
\end{equation}

G19 find that the amplitude of the halo response to the LMC is strengthened in Model 2 relative to the isotropic Model 1.
In this study, these two models will thus
allow us to explore whether the MW's initial halo kinematics also
 generate different kinematic responses in the Solar Neighborhood. 
For example, \citet{Green:2002a,Green:2002b} have suggested that 
halo velocity anisotropy can impact direct-detection experiments.
 
\begin{figure}[tbp]
\centering
\mbox{\includegraphics[width=4in]{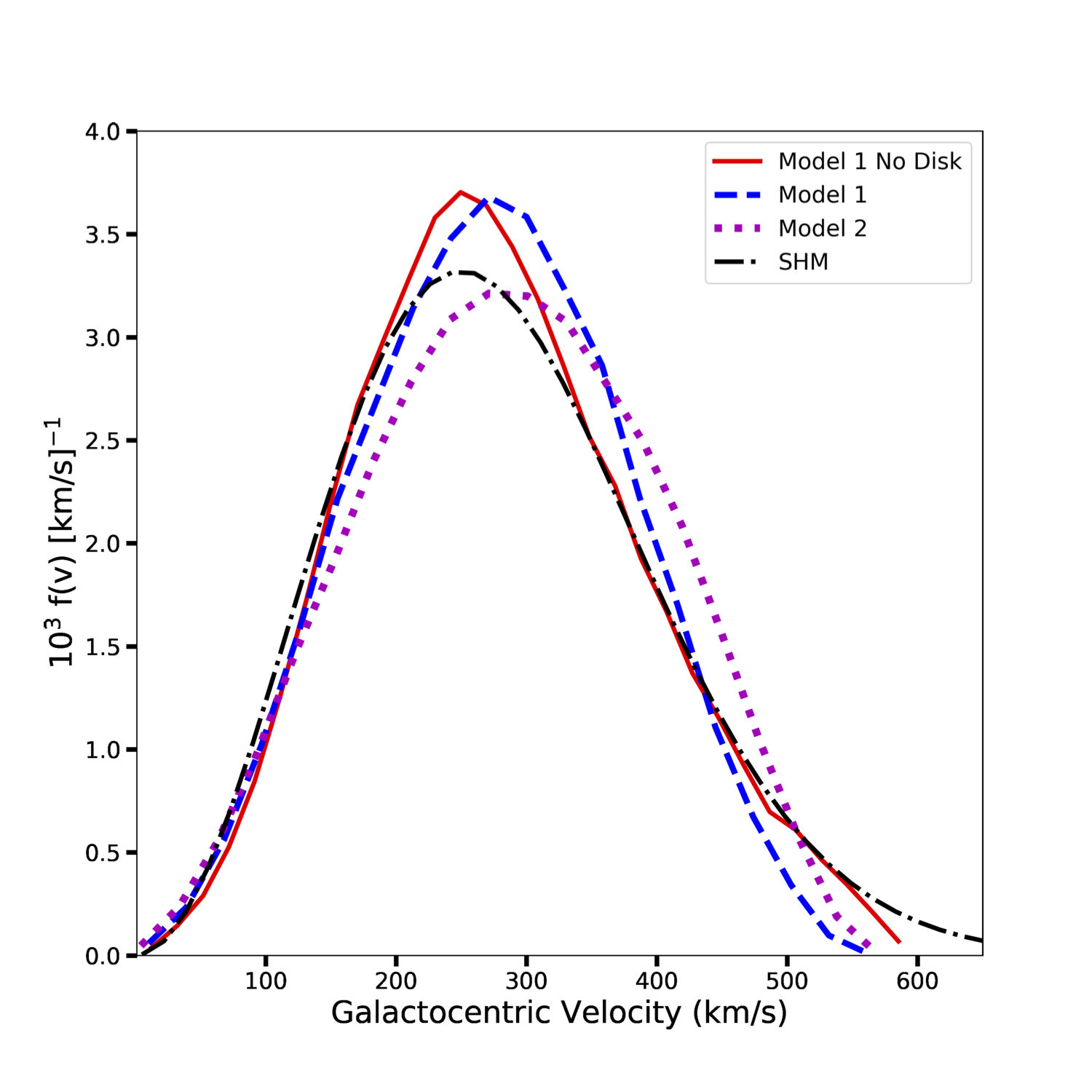}}
\caption{ Initial speed distributions for MW DM halo 
particles within a thick ring (width and thickness of 6 kpc), centered on the 
Solar radius of 8.3 kpc.  The MW is isolated; the LMC is not yet introduced.  The red, solid line 
indicates the distribution of halo particles in the isotropic Model 1, where a stellar disk 
and bulge is not included. This distribution roughly matches the 
corresponding SHM model (black, dash-dotted), computed with a 1D velocity 
dispersion of $\sigma = 177$ km/s. The inclusion of the disk and bulge in Model 1
shifts the distribution to higher speeds (blue, thick dashed line). When a radially 
biased $\beta(r)$ profile is adopted in Model 2 (purple, dotted line),
the discrepancy between the SHM model (which 
assumes isotropic dispersion) increases. 
In this study, we will compare perturbations induced by the LMC against these initial conditions 
for Model 1 and 2 (blue dashed and purple dotted lines). 
\label{fig:fv_initial}}
\end{figure}

Both the 
velocity anisotropy profile and the inclusion of a stellar disk/bulge
impact the kinematics of DM particles 
in the Solar Neighborhood.  This is illustrated in Figure~\ref{fig:fv_initial}, which 
shows the normalized distribution function of 
Galactocentric speeds, $f(v)$:  

\begin{equation}
    f(v) = \frac{dN}{dv}\frac{1}{N}
\end{equation}

\noindent of all DM particles
selected within a thick ring (6 kpc in both width and thickness),
centered at the Solar radius of 8.3 kpc. Results are shown for three different MW halos: Model 1, Model 2,
and Model 1 initialized without a disk or bulge. 

Plotted for reference is the 1D speed distribution 
function for the Standard Halo Model (SHM),
which has isotropic dispersions 
and is described by a Maxwell-Boltzmann distribution:

\begin{equation}
    f_{\rm SHM}(v) \propto e^{-(v^2/2\sigma^2)}v^2 dv \, \, \rm{ (km/s)}^{-1}
\end{equation}

The value of the 1D velocity dispersion ($\sigma = 177$ km/s) 
is computed from the 
motions of DM particles in the Solar Neighborhood
within Model 1, initialized without a disk. 
Note that the SHM speed distribution, $f_{\rm SHM}(v)$, is usually truncated 
at an escape speed of 500-600 km/s (e.g., as measured by the RAVE survey \cite{Piffl2014}), 
but we do not do this here to provide a point of reference for the high velocity tail.

The SHM model corresponds to a collisionless isothermal density distribution, with $\rho (r) \propto r^{-2}$ (see Ref. \cite{Peter2008}).
Unsurprisingly, the SHM model is not a good fit to the kinematics of the
DM halo in Model 1 (red solid line), 
which is modeled as a Hernquist density profile. 
In general, self-consistent, stable models of cuspy DM structures do not have a locally Maxwellian velocity distributions
\citep{Kazantzidis:2004}. 

Additionally, Figure~\ref{fig:fv_initial} illustrates that the
discrepancy between the SHM model and the modeled MW
increases when a stellar bulge and disk is included in Model 1 (blue dashed lined)
and when the halo kinematics are not isotropic, as in Model 2 (magenta dotted).

\subsection{The LMC Model}
\label{sec:LMC}

The LMC is modeled using a Hernquist DM halo of mass
\Mvir$ = 2.5 \times 10^{11}$ M$_\odot$. 
This is the highest LMC mass model explored in G19 (their model LMC4; see their table 2).
This LMC model is consistent with cosmological expectations 
of LMC analogs that host SMC-mass companions (see \S~\ref{sec:massivesat}). 
In \S~\ref{sec:LMCMass} we study the validity of our results in lower 
mass LMC models (G19 models LMC3 and LMC2). 

 The initial orbital parameters for the LMC are chosen such that its simulated present day 
 position and velocity are within 2$\sigma$ of the observed values 
 2 Gyr after first crossing the MW's virial radius~\citep{Kallivayalil:2013}.
 The LMC's orbital plane is roughly confined with the YZ Galactocentric plane, where the 
 LMC reached its pericentric approach of 48 kpc, 50 Myr ago.
 The fiducial simulations used in this study thus correspond 
 to Simulations 4 and 8 in G19 (see their table 3). 
 In the following we track 
 the present day positions and velocities of DM particles
 that were initially bound to the 
 LMC at infall but now reside within the Solar Neighborhood.

\subsection{Defining Motions in the Solar Neighborhood}
\label{sec:motions}

We identify particles representative of motions in the Solar Neighborhood 
as having Galactocentric positions: 

\begin{equation}
    |x- x_\odot |  < 3.0 \,\, {\rm kpc},
    \qquad
    |y- y_\odot |  < 3.0 \,\, {\rm kpc},
    \qquad
    | z- z_\odot |   < 3.0 \,\, {\rm kpc},
\end{equation}

\noindent where the position of the Sun is
($x_\odot, y_\odot, z_\odot$)=  (-8.3, 0, 0.027) kpc. 
There are $\sim 140,000$ MW DM particles in this volume in both Model 1 and Model 2
before the LMC crosses
the virial radius of the MW $\sim$ 2 Gyr ago. 
Note that because the flow of LMC particles is not uniform
throughout the disk, 
we do not consider the motions of particles within an annular ring centered 
at the Solar radius, in contrast to $\S$\ref{sec:initialkinematics}. 
The local MW DM density in this volume is $\sim$0.4 Gev/cm$^3$ for both Models 1 and 2.

We acknowledge that the
height of the defined Solar Neighborhood (3 kpc above or below the disk midplane) is large. This choice was
made to 
maximize the number of LMC particles in the volume, given the resolution of the G19 simulation.
On average, high speed LMC and MW DM
particles ( $>$ 800 km/s with 
respect to the Earth) are 
roughly evenly distributed at all distances above or below the disk midplane
(LMC high speed particles have a median height of $|z|$ $\sim$ 1.0 kpc from the disk midplane in the defined volume). 
As such, we are not averaging over local effects 
that would bias the expected velocity distribution of LMC particles in a smaller volume.

We compute the motion of LMC and MW particles in the Solar Neighborhood utilizing two different coordinate systems:

\begin{enumerate}
\item {\bf Galactocentric Coordinates:} A reference frame defined relative to the Galactic center. Because the MW and LMC are both 
modeled as live systems, the MW's center of mass (COM) will move in response to 
gravitational forces from the LMC \cite{Gomez:2015}.  We account for this motion by 
computing all particle velocities with respect to the MW COM. In this coordinate system, the 
LMC is currently moving predominantly in the -Y direction (opposite to 
the velocity vector of the Sun).

\item {\bf Earth Frame Coordinates:} A reference frame defined relative to the position of the Earth.   Particle velocities are corrected
for both the motion of the Local Standard of Rest (LSR)\footnote{Our
results are insensitive to the exact value of the LSR. The LSR motion is removed from the LMC's observed velocity vector, measured with respect to the Sun \citep{Kallivayalil:2013}, in order to compute the simulations in a Galactocentric frame. However, we then add back the LSR motion to carry out the calculations presented in this work. 
} and the peculiar motion of the Sun.
These are defined in Galactocentric coordinates following \citet{Schonrich:2010} and \citet{McMillan:2017}:
\begin{equation}
    Vx_\odot = 11.1 \, \, {\rm km/s,}
    \qquad 
    Vy_\odot = 12.24 + 239.0 \, \,{\rm km/s,}
    \qquad 
    Vz_\odot = 7.25\, \,{\rm km/s}.
\end{equation}

To account for the annular motions of the Earth 
about the Sun, we follow equation 4.9 and 4.10 
in Ref. \cite{Kelso:2016} (see also Refs. \cite{Green:2002a,Freese:2013}).

Results presented in the Earth Frame thus account 
for both the motion of the Local Standard of Rest, the peculiar motion of the Sun and 
the annular motions of the Earth averaged over the year, unless 
otherwise indicated (e.g., at the end of \S\ref{sec:results} and \S\ref{sec:direct}).
Because the LMC is currently moving in the
opposite direction of the Sun, its DM 
particles will flow through the Solar Neighborhood 
in roughly the same direction as that of 
the MW halo particles.

\end{enumerate}

\section{Results: DM Motions in the Vicinity of the Sun}
\label{sec:results}

\begin{figure}[tbp]
\centering
\mbox{\includegraphics[width=3in, trim={0.3in 0.5in 0.6in 0.6in}, clip]{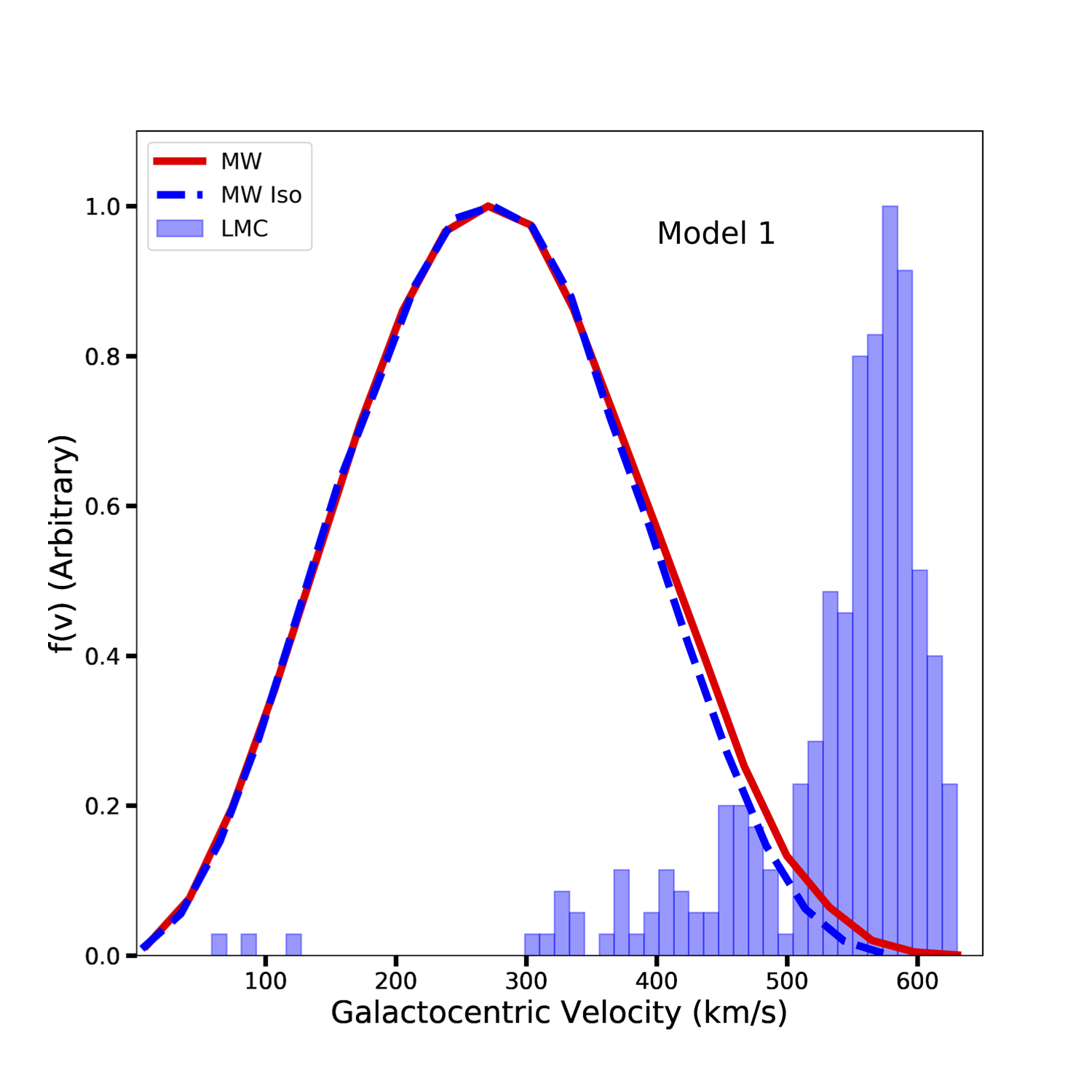}
\includegraphics[width=3in, trim={0.3in 0.5in 0.6in 0.6in}, clip]{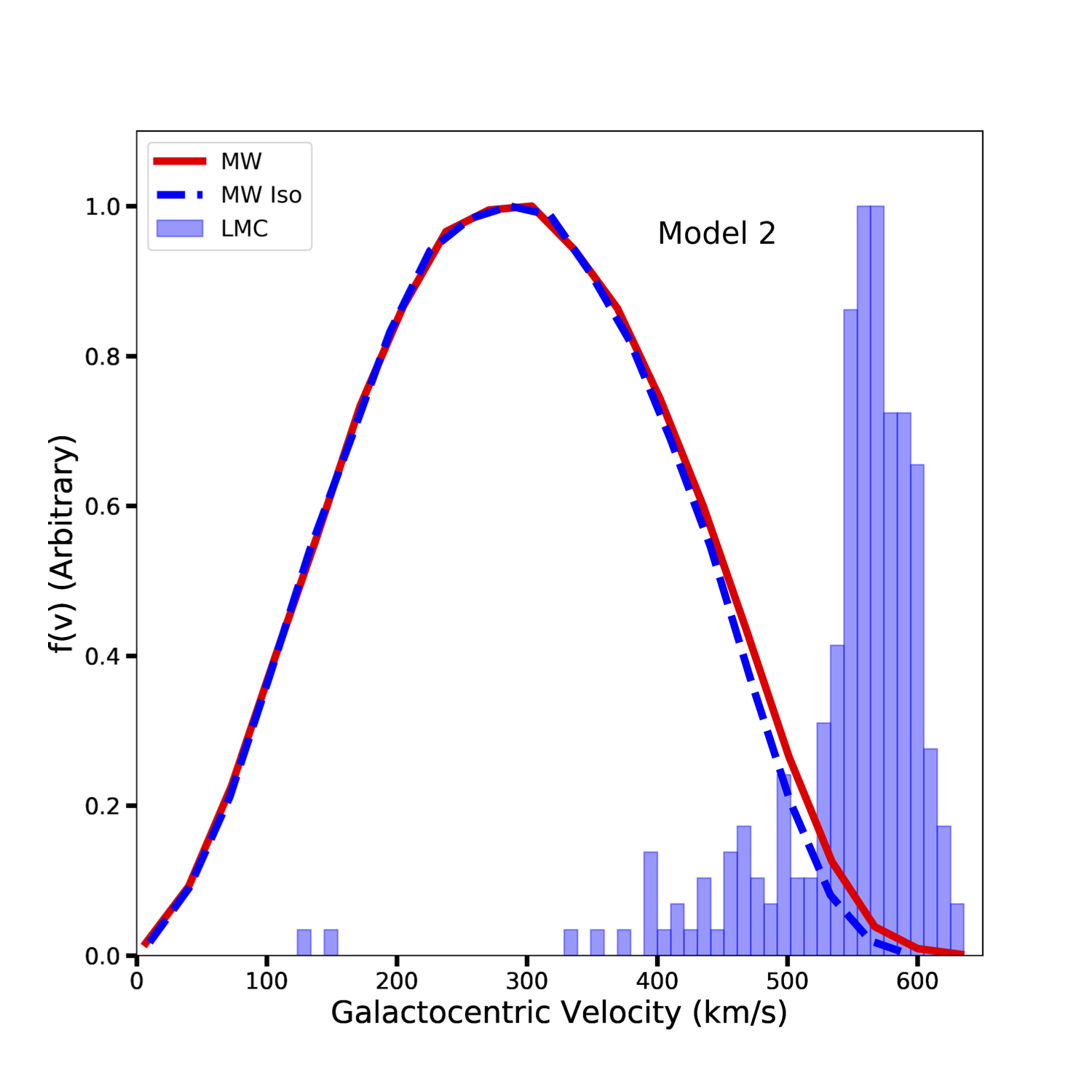}}
\caption{Galactocentric velocity distribution for all DM particles in the 
Solar Neighborhood, as defined in \S~\ref{sec:motions}. Each distribution is arbitrarily 
normalized. LMC particles contribute at high speeds, primarily from 550 km/s$-$640 km/s (blue histograms), with a median value of 570 km/s in both Model 1 (left panel) and Model 2 
(right panel).  These speeds are near or exceed the local escape speed. The initial speed distribution of MW halo particles, 
before the LMC is introduced 
(blue dashed line) is skewed to higher speeds after the LMC arrives at its current location
(red solid line) -- no LMC particles are included in these MW distributions. 
Results for Model 1 (left) and Model 2 (right) are largely
similar, with Model 2 generally having more particles at the highest speeds.  
\label{fig:Galactocentric} }
\end{figure}

In Figure~\ref{fig:Galactocentric}, we show the DM speed distribution in the Solar Neighborhood, with the inclusion of the LMC, in Galactocentric coordinates.  We illustrate the speed distributions of MW DM particles for the two different MW models, 
before (blue dashed line) and after (red solid line and 
blue histogram) the 
LMC is introduced. DM particles from the LMC alone (not MW particles perturbed by the LMC's passage through the disk) are shown with the histogram. Each distribution is normalized to peak at 1.0.

The speed distribution of LMC particles is striking. The RAVE survey finds that the 90\% confidence range for the local escape 
velocity is 492 km/s$-$587 km/s in the Galactocentric frame \cite{Piffl2014}. However, 
LMC particles are moving predominantly at 550-640 km/s --- higher than, or at the same speed as, the
local escape speed.
Because the LMC is coming in for the first time, its contributing particles are not 
in equilibrium and can be moving at high speeds. Comparing the two panels of Figure~\ref{fig:Galactocentric}, we find that the resulting LMC particle kinematics are the 
same in both Model 1 and Model 2 (median velocity of $\sim$570 km/s). This 
is by design as the LMC mass model and orbit is the same in both cases.

While LMC particles comprise a small fraction ($\sim$0.2\%) of particles in the Solar Neighborhood,
they dominate the high speed tail - 70\% of DM particles with geocentric speeds $>750$ km/s originate from the LMC.
Furthermore, the LMC also impacts the kinematics of MW halo particles, an effect which extends beyond the Solar Neighborhood, affecting stars as well as DM in the MW halo \cite{G19}.
As the LMC passes near the MW disk, it accelerates DM particles to higher speed, such that the {\it local} MW DM velocity distribution 
skews to higher speeds (red solid line in Figure~\ref{fig:Galactocentric}). 20\% of DM particles moving 
at geocentric speeds $>600$ km/s either originate from or are accelerated by the LMC, together 
augmenting the local DM density distribution by $\sim$5\%.

\begin{figure}[tbp]
\centering
\mbox{\includegraphics[width=3in, trim={0in 0.5in 0.6in 0.6in}, clip]{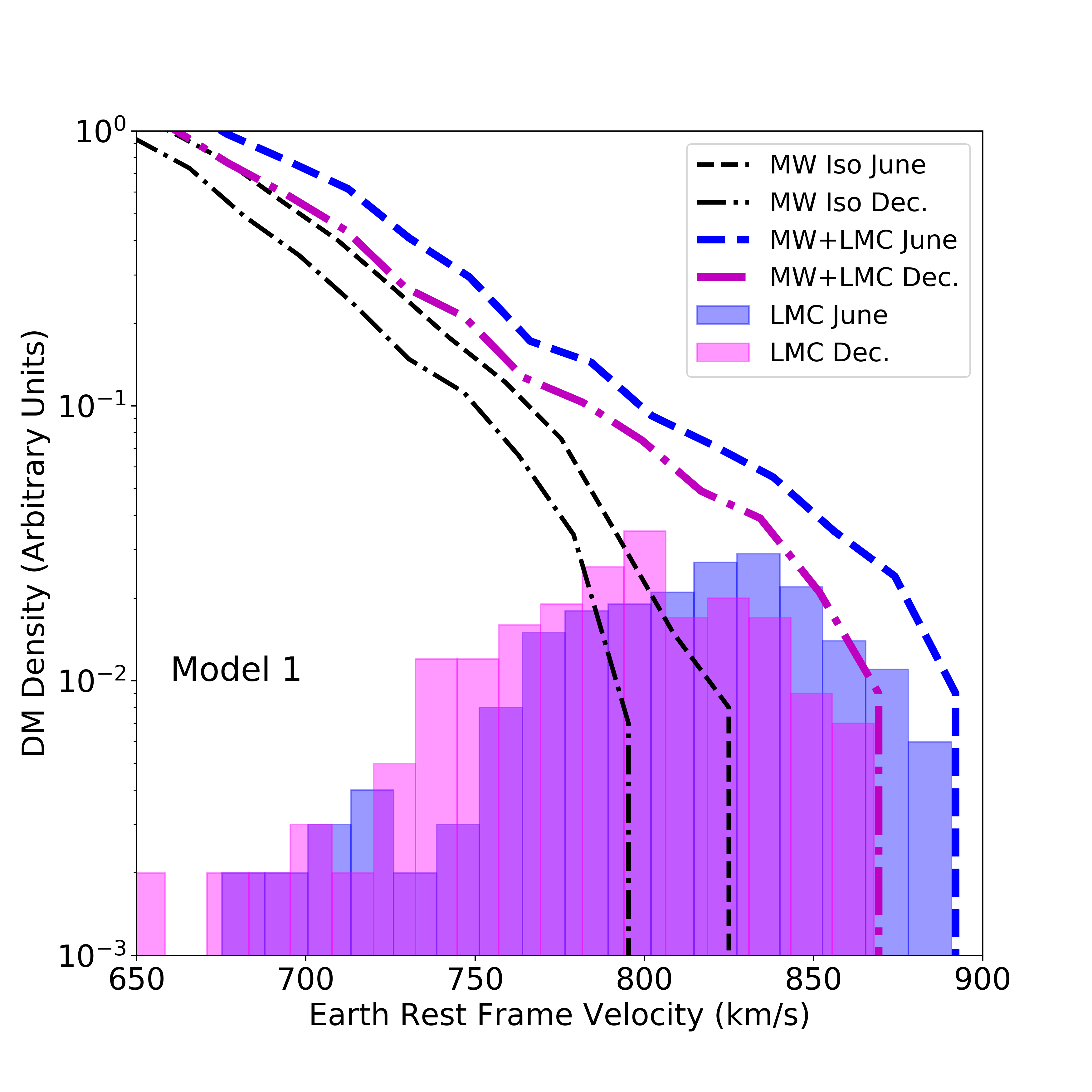}
\includegraphics[width=3in, trim={0in 0.5in 0.6in 0.6in}, clip]{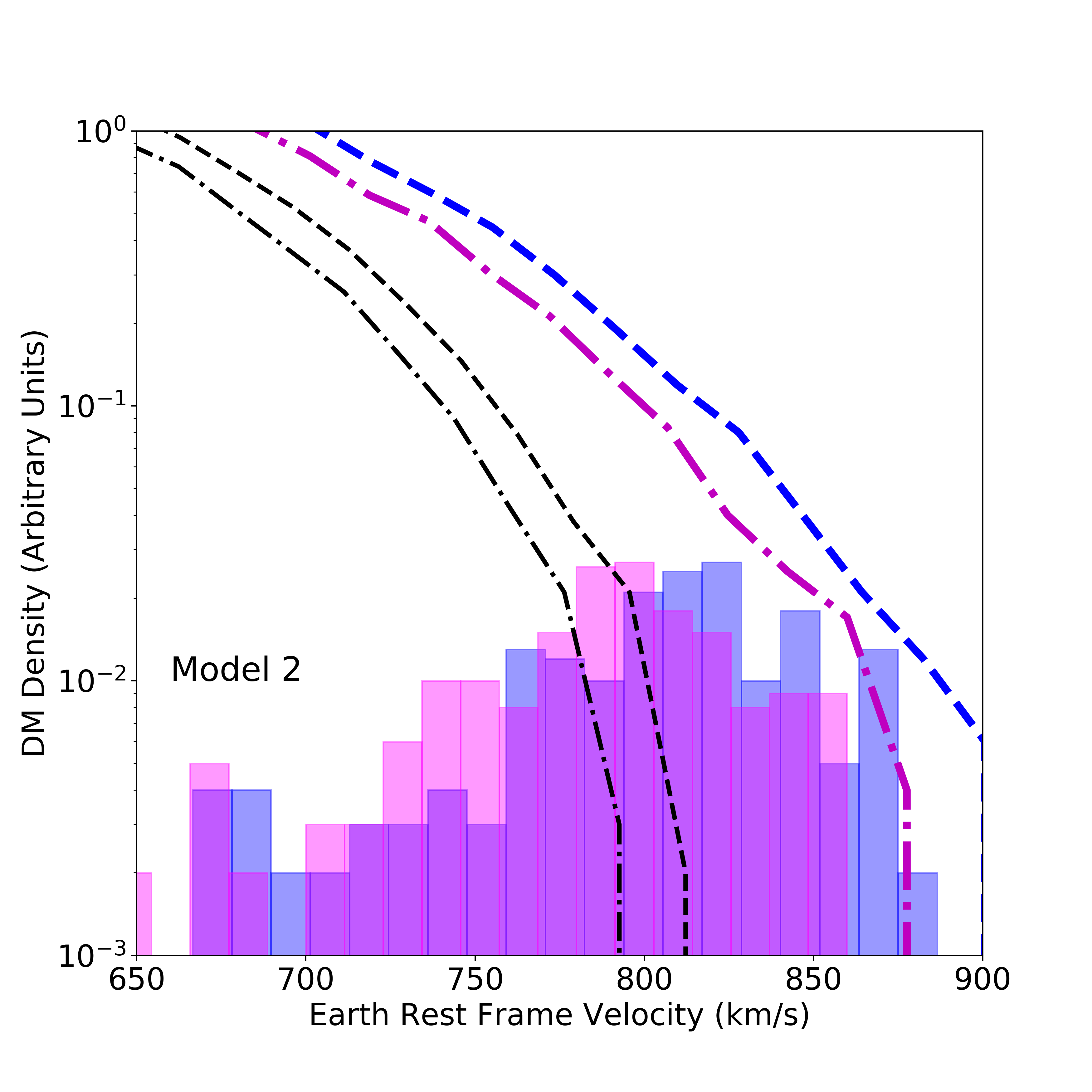}}
\caption{Speed distribution in Earth Frame Coordinates for all DM particles in the 
Solar Neighborhood for Model 1 (left) and Model 2 (right). The Y-axis indicates the relative density of DM 
particles in the solar volume, in arbitrary units. 
Thin black lines represent the initial high-speed Earth Frame speed 
distribution of MW DM particles (before the LMC arrives; MW Iso).
Dashed and blue lines indicate the 
speeds of MW and LMC DM particles expected in June (maximal) and dashed-dotted and magenta 
lines indicate speeds in December (minimum).  DM particles that originate in the LMC are also plotted separately as histograms, following the same color coding. In December(June), all particles moving at 
speeds in excess of $\sim$800(840) km/s either originate from or are accelerated by the LMC. 
\label{fig:Earth} }
\end{figure}

The relative contribution of LMC particles to the high-speed tail compared to the acceleration of MW particles by the LMC is illustrated more clearly in the Earth Frame Coordinate system (\S\ref{sec:motions}) in Figure~\ref{fig:Earth}.  In that Figure, we focus on geocentric speeds greater than 650 km/s.  The inclusion of the LMC clearly skews the velocity distribution 
to higher velocities, boosting particle speeds to as high as 900 km/s in June 
and 875 km/s in December. We will return to this point in the discussion of annual modulation of direct-detection signals in \S\ref{sec:direct}. Such speeds are not 
exhibited by MW particles prior to the arrival of the LMC (lines marked as MW Iso). 
As such, all of the highest speed particles in the Solar Neighborhood (near or above the escape speed) either originate from 
the LMC (60-100\%) or are MW particles that have been accelerated to high speeds by the LMC. This statement follows from the LMC's unique status 
as the only known substructure on first infall 
that can also impact the Solar Neighborhood, enabling DM particles to reach speeds near or exceeding the local escape speed.

 Table~\ref{table:highspeed} lists the DM density in the high speed
 tail originating from either the LMC or the MW (Model 1 or 2, before, MW$_{Iso}$, and after the 
 LMC is accreted; columns 3-6).   The inclusion of the LMC 
increases the amount of DM at Earth Frame velocities greater than 800 km/s by factors of 4-11 relative to the MW in isolation and 
by factors of 2 in June vs. December (column 7). For Model 1, the highest-speed particles come from the LMC, but for Model 2, the high-speed population has a higher contribution from accelerated MW halo particles (column 8). This result is consistent with the stronger halo response to the LMC's passage found by G19 in Model 2 -- i.e. the density enhancements in the Transient and Collective Responses in Model 2 are larger than in Model 1. Finally, all DM particles moving at geocentric speeds higher than 850 km/s either originate in or are accelerated by the LMC.

\begin{table*}
\centering
\caption{ {\bf Density of High Speed DM Particles in the Solar Neighborhood (10$^3$M$_\odot$/kpc$^3$)}
\newline
MW$_{\rm Iso}$ refers to the MW before the LMC arrives.}
\label{table:highspeed}
\begin{tabular}{ |c|c|c|c|c|c|c|c| }
\hline\hline
Model & Speed (km/s) &  MW$_{\rm Iso}$ & MW & LMC  & MW+LMC & $\frac{\rm{MW+LMC}}{\rm{MW}_{\rm Iso}}$ & $\frac{{\rm LMC}}{{\rm MW+LMC}}$\\
\hline
1  & V$_{\rm June}$ $>800$  & 1.9  & 8.3 & 8.3 & 16.7 & 9.0 &  0.50\\
1  & V$_{\rm Dec}$ $>800$   & 0.93  & 3.2 & 6.5 & 9.3 & 11.0  & 0.35 \\
\hline
1  & V$_{\rm June}$ $>850$   & 0.0  & 0.93 & 3.2 & 4.2 & - & 0.77  \\
1  & V$_{\rm Dec}$ $>850$   & 0.0  & 0.0 & 0.93 & 0.93 & - &  1.0 \\
\hline
2  & V$_{\rm June}$ $>800$  & 4.7 & 13.0 & 6.9 & 20.0 & 4.3 & 0.35\\  
2  & V$_{\rm Dec}$ $>800$   & 1.9 & 6.0 & 4.7 & 10.6 & 5.8 & 0.43\\
\hline
2  & V$_{\rm June}$ $>850$  & 0.0 & 1.4 & 2.3 & 3.7 & - & 0.63 \\ 
2  & V$_{\rm Dec}$ $>850$   & 0.0 & 0.46 & 0.93 & 1.4 & - & 0.67 \\ 

\hline\hline
\end{tabular}
\end{table*}

\begin{figure}[tbp]
\centering
\mbox{\includegraphics[width=5in, trim={1.5in 4in 1.5in 4in}, clip]{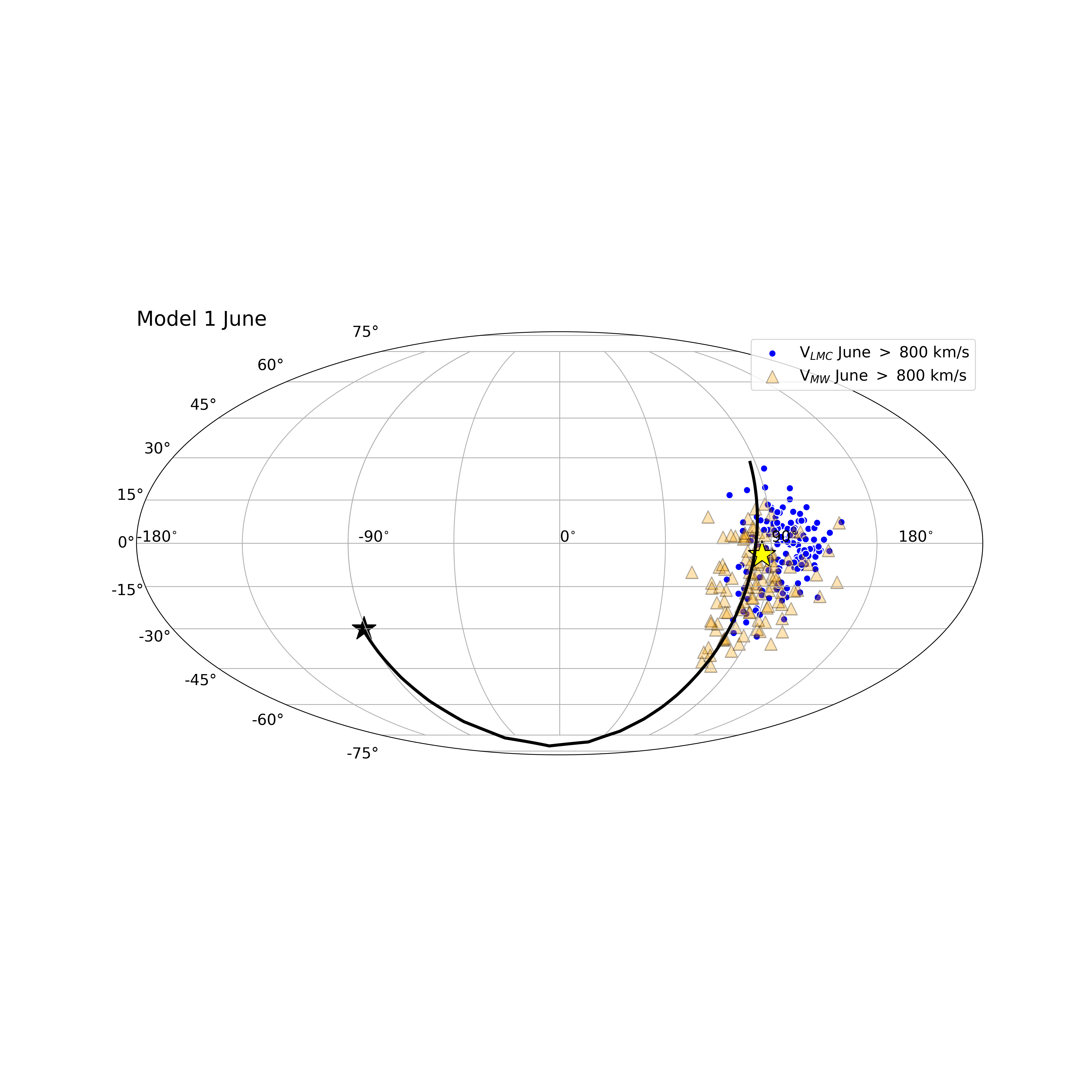}}
\mbox{
\includegraphics[width=5in, trim={1.5in 4.3in 1.5in 4in}, clip]{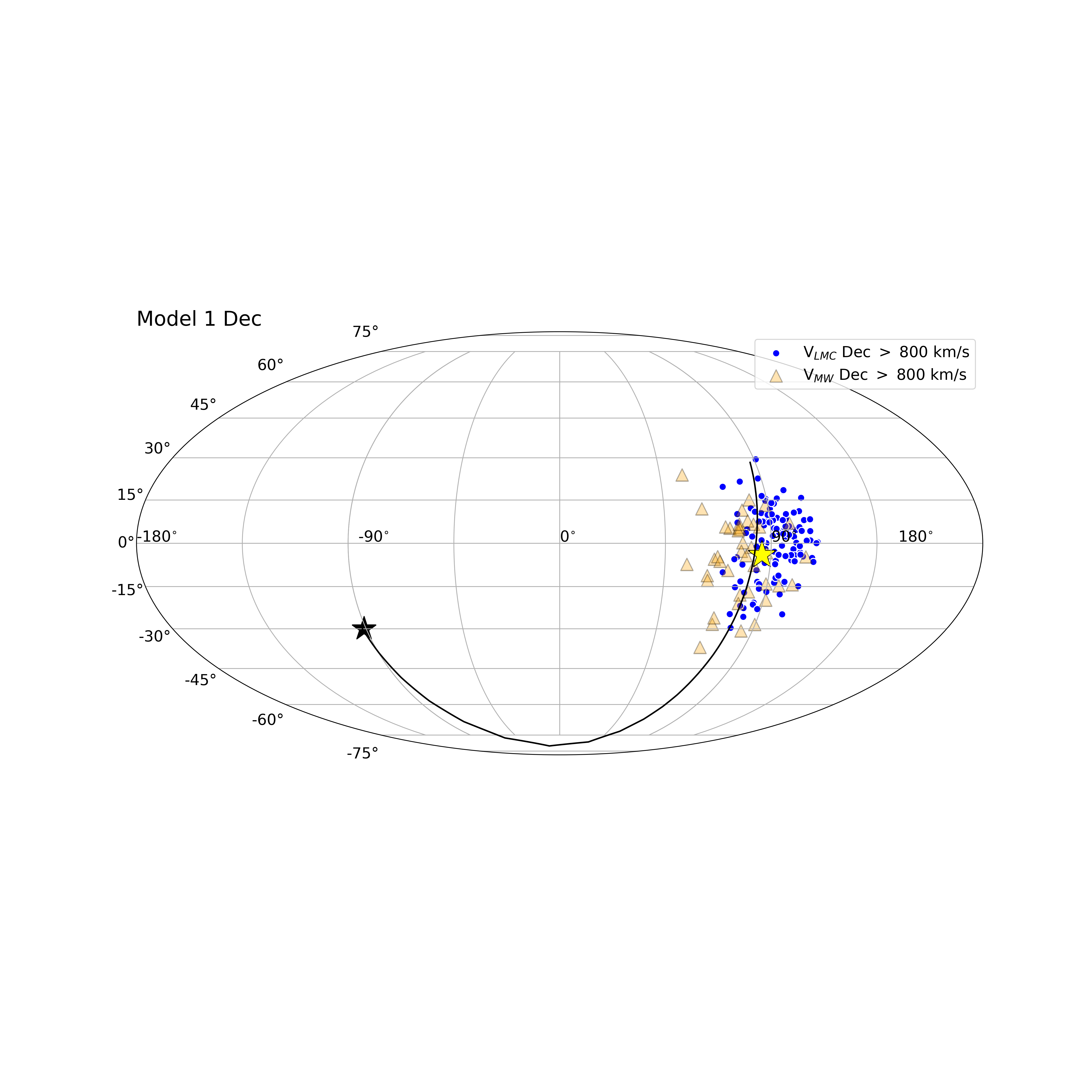}}
\caption{Mollweide all-sky, equal-area, projection of the trajectories of all DM
particles in the Solar Neighborhood for Model 1 in Earth Frame 
Coordinates in June (top) and December (bottom). Trajectories indicate the direction 
from which the particles originate. 
The current location of the LMC (black star) 
and its orbital path over the past 2 Gyr (black line) are marked.
The reflex motion of the Sun is indicated by the yellow star, which is 
coincident with the direction of motion of the LMC at closest approach, and hence also of the trajectories of both DM particles that originate within (blue circles) or are accelerated by (orange triangles) the LMC. 
\label{fig:Mollweide1} }
\end{figure}

We can also show the direction from which the particles entered the Solar Neighborhood in the Earth Frame Coordinate system, which is of interest for directionally sensitive experiments \cite{Mayet:2016zxu}.  Figure~\ref{fig:Mollweide1} illustrates the trajectories of 
high-speed DM particles in the Solar Neighborhood  ($v >$ 800 km/s in Earth Frame Coordinates) projected on Mollweide equal-area plots of the sky at two different times of year.  Trajectories are computed 
as the negative of the Earth Frame velocity vector of each DM particle, thus indicating the direction from which they originate. 
Results are presented for MW Model 1 and are similar for Model 2.

 Particles at these high speeds 
 primarily originate within or are accelerated by the LMC (Table~\ref{table:highspeed}), as such their motions reflect 
 the direction of motion of the LMC at closest approach. This direction is coincident with the reflex motion of the Sun, indicated by the yellow star, explaining why the  
LMC has such a dramatic effect on the local velocity distribution at high speeds. If we attempted this 
experiment on the opposite side of the MW's disk, where the LMC's motion would be opposite the reflex motion 
of the Sun, the speeds would be roughly 240 km/s lower than we find here.  

Although the LMC boosts the density of high-speed DM in the Solar Neighborhood, the arrival direction is approximately the same as (although somewhat more southerly than) the highest speed particles in the SHM.  The arrival direction is little changed between the times of maximal (June) and minimal (December) relative speed of the Earth with respect to the Galactic halo, although the density of particles is higher in the former case (top vs. bottom panels in Figure~\ref{fig:Mollweide1}).  Interestingly, the arrival direction is distinct from other kinematic stellar substructures observed in Gaia \citep{Evans:2019bqy}.

\section{Implications for direct-detection limits}\label{sec:direct}

We consider the implications of the LMC's presence in the MW for WIMP direct-detection specifically, although the results from \S~\ref{sec:results} may also be applied to direct searches for other DM candidates.  For concreteness, we consider only spin-independent elastic scattering off nuclei, although our results generalize to other types of interaction with nuclei or electrons.

For spin-independent interactions, the expected differential event rate is

    \begin{eqnarray}\label{eq:energyspectrum}
  \frac{dR(Q,t)}{dQ} = \left( \frac{\rho_\chi}{m_\chi} \right) \epsilon(Q) 
  \sum_A N_A \int \limits_{v_{\mathrm{min}(m_\chi,m_A,Q)}} \, d^3 v \frac{d\sigma_A(v)}{dQ} |\mathbf{v}| f(\mathbf{v},t),
\end{eqnarray}
the number of events in the experiment per unit time $t$ per unit recoil energy $Q$. We assume that the local phase-space density can be represented by the multiplication of the DM particle number density (the local mass density $\rho_\chi$ divided by the WIMP particle mass $m_\chi$) and utilize the three-dimensional velocity distribution $f(\mathbf{v},t)$ in Earth Frame Coordinates.  We assume that the experiment has a set of isotopes $\{A\}$, and that there are $N_A$ target nuclei of isotope $A$ in the experiment, each with mass $m_A$.  The efficiency for an experiment to detect nuclear recoils of energy $Q$ is $\epsilon(Q)$. For a WIMP to scatter a nucleus to energy $Q$, the WIMP must have a minimum speed
\begin{eqnarray}
  v_{\mathrm{min}}  = \sqrt{\frac{m_A Q}{2 \mu_A^2}},
\end{eqnarray}
where $\mu_A$ is the reduced mass of the nucleus-WIMP system.  The elastic scattering cross section is 

\begin{eqnarray}\label{eq:sisdsigma}
  \frac{d\sigma_{\mathrm{A}}}{dQ} = \frac{m_A}{2\mu_A^2 v^2}\left[ \sigma_{\mathrm{A}}^{\mathrm{SI}} F^2_{\mathrm{SI}} (Q) \right],
\end{eqnarray}
where $\sigma^{\mathrm{SI}}_{\mathrm{A}}$ is the interaction cross section in the limit of no momentum transfer, and $F(Q)$ is the form factor (taken to be the Helm form factor \cite{Engel:1991wq} in this work).  

In order to benchmark experiments against each other, it is useful to express Eq. (\ref{eq:energyspectrum}) in terms of the WIMP-proton cross section \sigmapsi~and to isolate the universal velocity dependence of the integral from other variables that depend on experimental specifics or WIMP mass.  If the WIMP interacts equally strongly with protons and neutrons, then
\begin{eqnarray}
  \sigma_{\mathrm{A}}^{\mathrm{SI}} &=& \left( \frac{\mu_A}{\mu_p} \right)^2 A^2 \sigmapsie,
\end{eqnarray}
where $\mu_p$ is the reduced mass of the WIMP-proton system and $A$ is the atomic number of the isotope.  The mean inverse speed $g(v_{\rm min})$ is
\begin{eqnarray}
  g(v_{\rm min}) = \int \limits_{v_{\mathrm{\rm min}(m_\chi,m_A,Q)}} \, d^3 v  f(\mathbf{v},t) / |\mathbf{v}|, 
\end{eqnarray}
and hence, Eq. (\ref{eq:sisdsigma}) can be re-written as
  \begin{eqnarray}
  \frac{dR(Q,t)}{dQ} = \left( \frac{\rho_\chi}{m_\chi} \right) \epsilon(Q) 
  \sum_A N_A \left( \frac{\mu_A}{\mu_p} \right)^2 A^2 \sigmapsie \frac{m_A}{2\mu_A^2} F^2(Q) g(v_{\mathrm{min}}).
\end{eqnarray}

\begin{figure}[tbp]
\centering
\mbox{\includegraphics[width=3.3in, trim={0in 0.5in 0.7in 0.5in}, clip]
{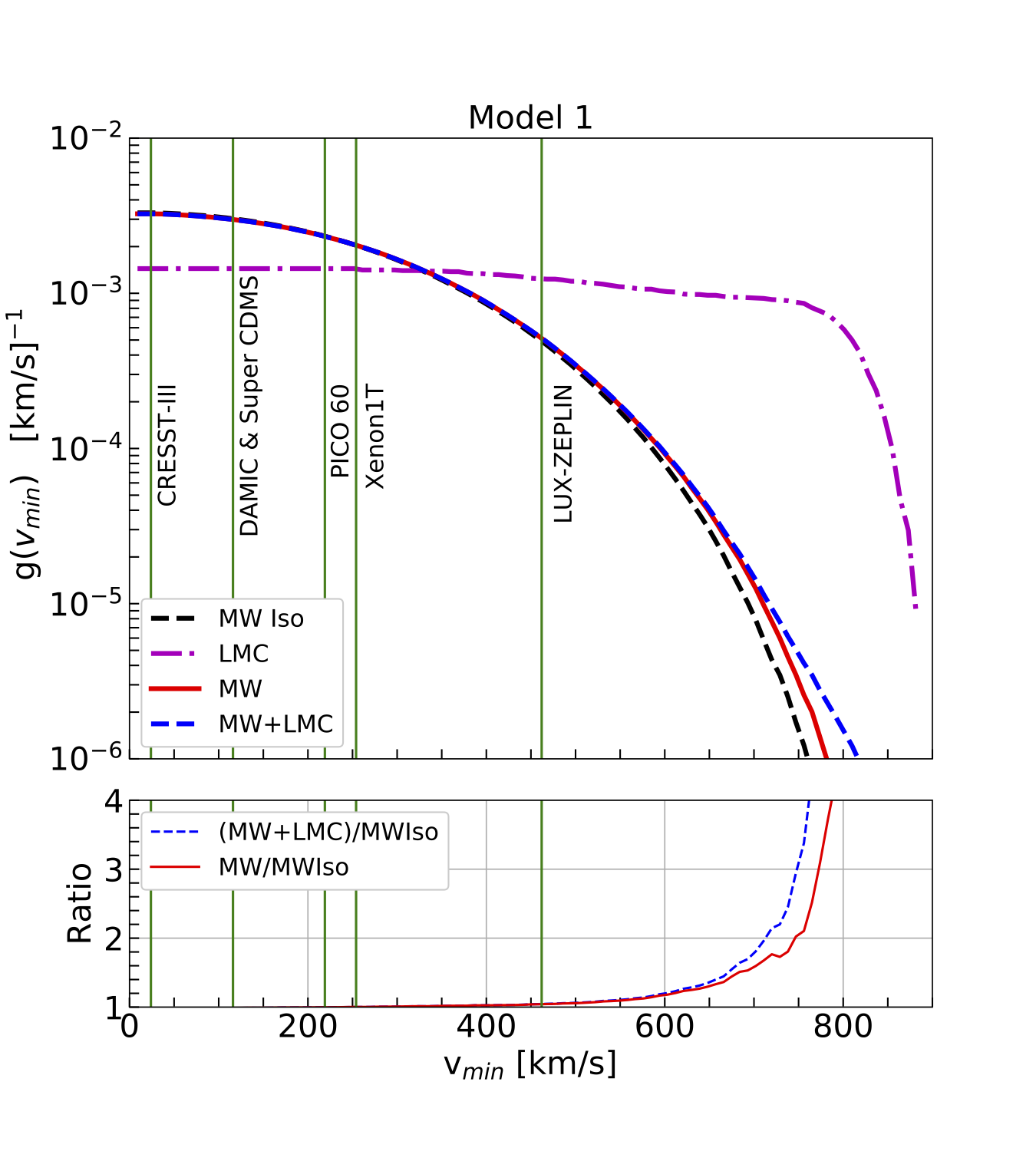}
\includegraphics[width=2.9in, trim={0in 0.5in 0.7in 0in}, clip]{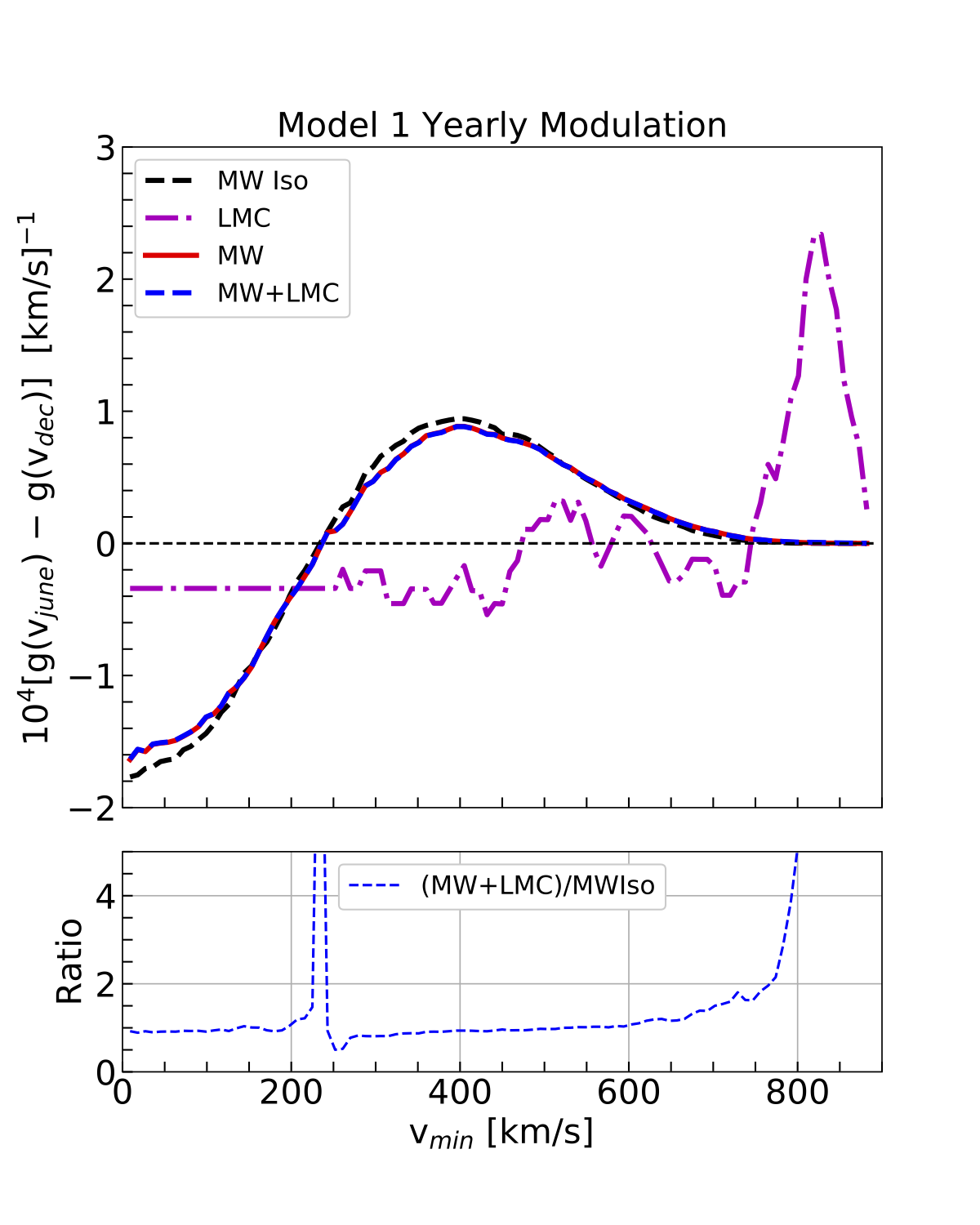}}
\caption{ {\it Left (top)}: Time-averaged inverse speed, $g(v_{\rm min}$), as a function of the minimum scattering speed $v_{\rm min}$ for halo Model 1 (isotropic). 
The black, dashed line indicates results for the Isolated MW (MW Iso), before the LMC is not introduced. Remaining lines indicate results after the LMC is introduced: the LMC particles 
only (LMC: purple, dashed-dotted); the MW particles only (MW: red, solid); and 
both the LMC and MW particles together (MW+LMC: blue, dashed).
Vertical green lines represent the value of $v_{\rm min}$ near
the thresholds for particle scattering of a 10 GeV/c$^2$ DM particle 
in the: Xenon1T ($Q_{min}$=1 keV, Xenon target) \cite{Aprile:2018}, 
LUX-ZEPLIN ($Q_{min}$=3.3 keV, Xenon target)\cite{Akerib:2018,Akerib:2016},
PICO 60 ($Q_{min}$=2.45 keV, Fluorine target) \cite{Amole:2019},
DAMIC ($Q_{min}$=0.6 keV, Silicon target) \cite{Aguilar-Arevalo2016}, 
SuperCDMS ($Q_{min} = 0.27$ keV, Germanium target) \cite{Agnese:2017}, and
CRESST-III ($Q_{min}$=0.03 keV, Oxygen target) \cite{Abdelhameed:2019} detectors.
{\it Left (bottom)}: The ratio of mean inverse speeds for MW particles only relative to the Isolated MW (red) and the ratio of the MW + LMC particles relative to the Isolated MW (blue, dashed). 
{\it Right (top)}: Expected annual modulation amplitude between June 
December as a function of $v_{\rm min}$. Line types are the same as in the left figure.  {\it Right (bottom)}: The ratio of results for the MW halo particles and the MW+LMC particles relative to the MW in isolation. The feature at $\sim$220 km/s is an artifact resulting
from a minor shift in the zero point between the two curves. 
}
\label{fig:gvminModel1}
\end{figure}

\begin{figure}[tbp]
\centering
\mbox{\includegraphics[width=3.3in, trim={0in 0.5in 0.7in 0.5in}, clip]{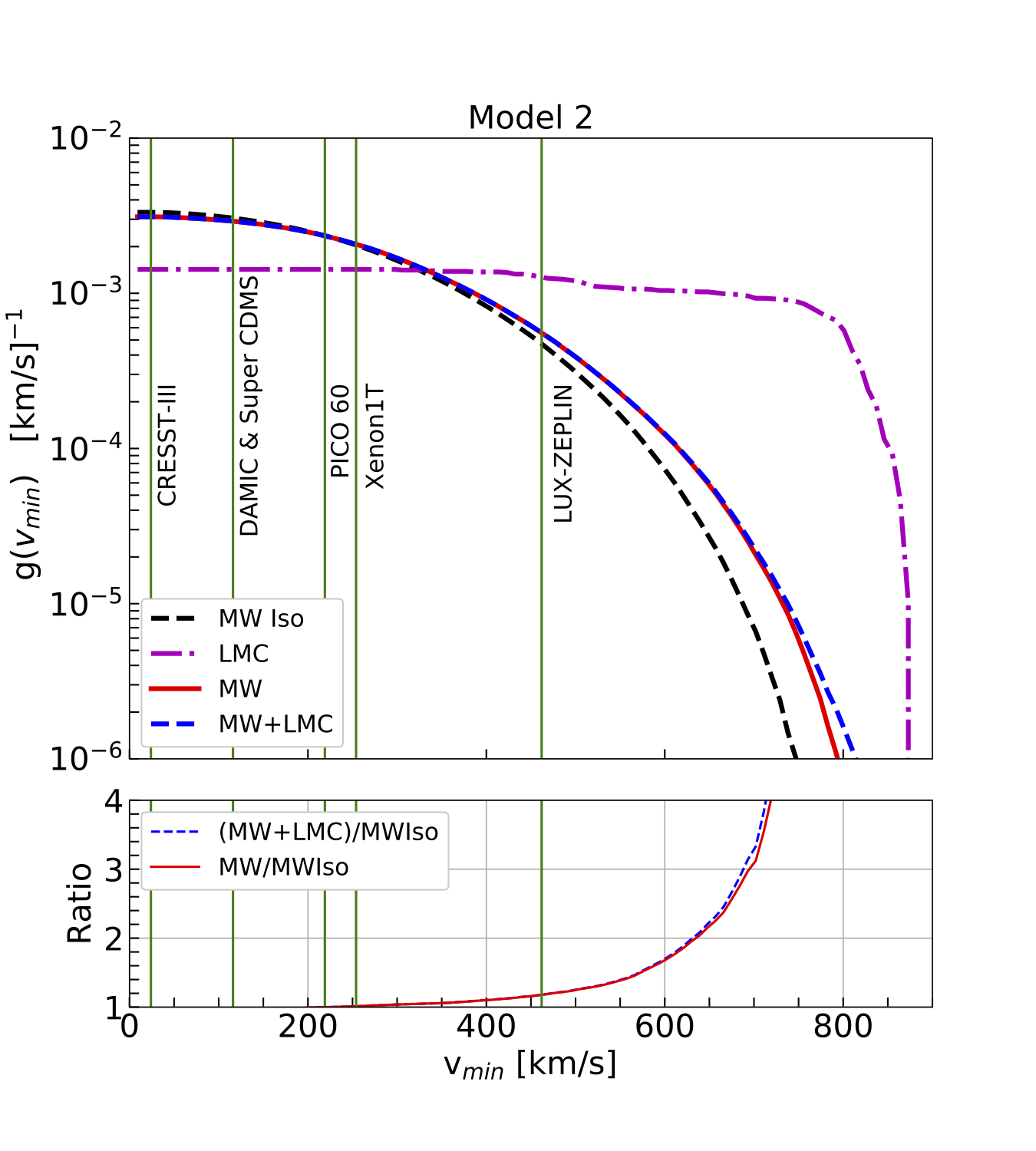}
\includegraphics[width=2.9in, trim={0in 0.5in 0.7in 0in}, clip]{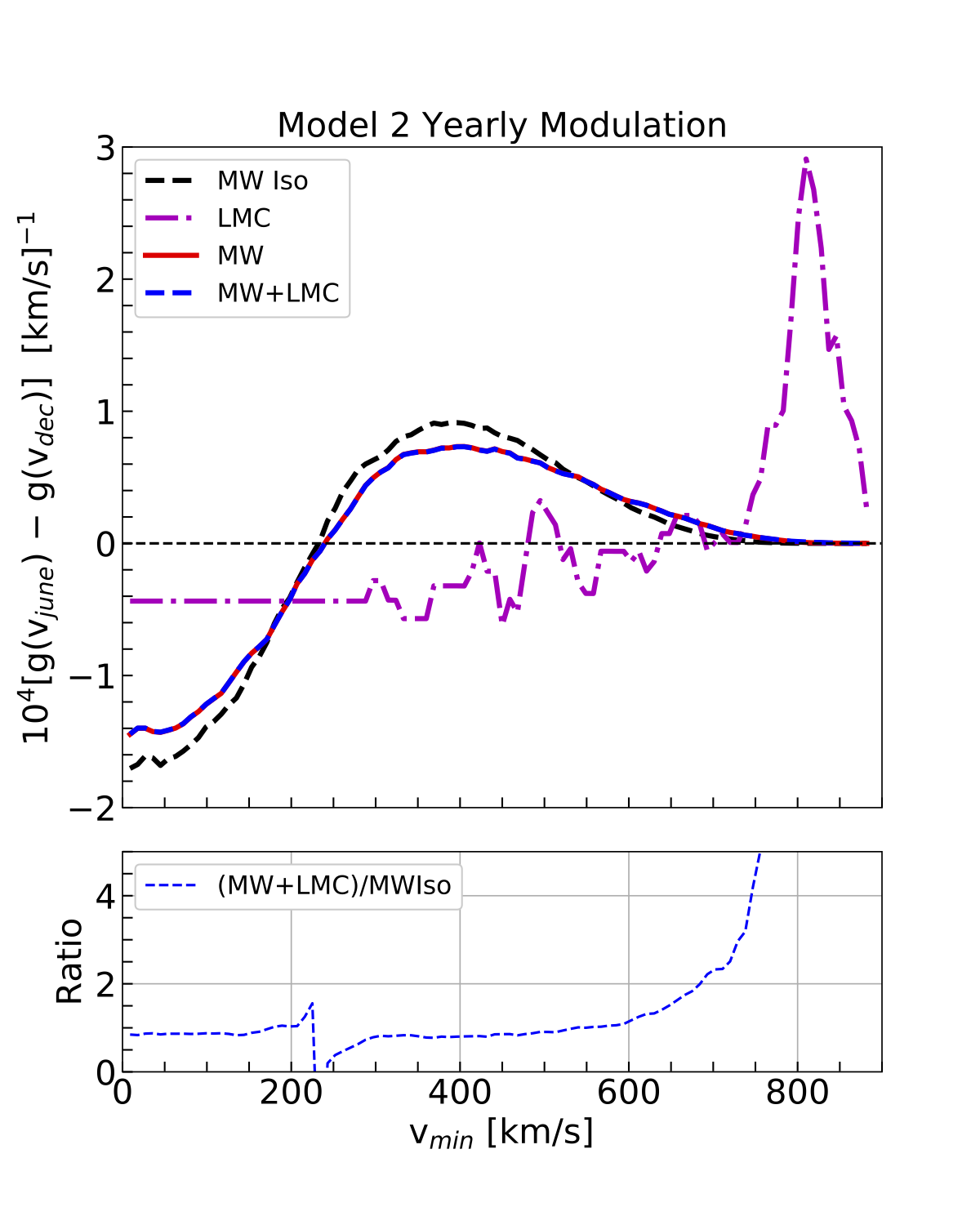}}
\caption{ Same as Figure~\ref{fig:gvminModel1}, but for Model 2 (anisotropic MW halo). The ratio of $g(v_{\rm min})$ for the perturbed MW and LMC relative to the Isolated MW (no LMC) indicate a larger increase in the event rate relative to Figure~\ref{fig:gvminModel1}, on account of the significantly higher acceleration of MW particles by the LMC in the anisotropic MW model relative to the isotropic model.}
\label{fig:gvminModel2}
\end{figure}

Modulo the form-factor suppression represented by $F^2(Q)$, the energy spectrum of events follows the mean inverse speed function $g(v_{\rm min})$. In Figures~\ref{fig:gvminModel1} \& \ref{fig:gvminModel2}, we show how the presence of the LMC alters $g(v_{\rm min})$, averaged over a year. The MW's mean inverse speed before the LMC is added is denoted by the dashed black line, labeled ``MW Iso''.  To separate the effects of the acceleration of MW particles by the LMC (red line) from particles of LMC origin, we plot both the mean inverse speed of LMC particles (dot-dashed purple) and of the MW + LMC system (dashed blue line).  The 3D velocity distribution function for each line is normalized to unity; in practice, the total event rate scales with the dark matter density $\rho_{\chi, i}$ in each component $i$. In the bottom-left panel, we show the ratio of the MW mean inverse speed to that of the MW Iso model, and the ratio of the mean inverse speed of all particles in our simulated Solar Neighborhood model to the MW Iso initial conditions.  For reference, we show the $v_{\rm min}$ corresponding to the minimum nuclear recoil energy thresholds for various experiments assuming a 10 GeV WIMP, taking the threshold explicitly indicated in papers or the recoil energy corresponding to a detection efficiency~$\epsilon(Q_{min}) = 0.5$.

The LMC is found to increase the 
$g(v_{\rm min})$ both by accelerating MW particles and also by contributing 
high-speed DM particles. Because the LMC particles are dominantly near the escape velocity from the MW, the mean inverse speed is high (dot-dashed line), and the energy spectrum is hard.   However, the percentage of Solar Neighborhood particles originating in the LMC is relatively small, and the bigger effect of the LMC on the mean inverse speed function is the acceleration of MW particles by resonances induced in the halo by the LMC, until the LMC becomes dominant for speeds above 750 km/s.  This is most visible in the bottom left panels of Figures~\ref{fig:gvminModel1} \& \ref{fig:gvminModel2}, where we examine the ratio of $g(v_{\rm min})$ of Solar Neighborhood particles (either with or without the LMC particles) after the LMC's passage relative to the isolated MW.  Above a geocentric speed of 600 km/s (close to the threshold for S1 + S2 analyses of xenon nuclear recoil analysis for a 10 GeV WIMP), the LMC's passage leads to a 25\% enhancement of the signal in Model 1, and nearly a factor of 2 change for Model 2.  Especially for Model 2, the MW response to the LMC passage leads to a large enhancement in $g(v_{\rm min})$, which increases sharply to factors of a few beyond $v_{\rm min} = 600$ km/s.  Above the escape speed from the MW ($\approx$ 750-800 km/s in Earth Frame Coordinates; Figure~\ref{fig:Earth}), the ratio is undefined, and the LMC dominates the signal.

Before we describe the key effect of the LMC on WIMP direct-detection -- the tightening of exclusion curves at low WIMP mass -- we highlight the effect of the LMC on the direct-detection annual modulation signal.  In the right-hand panels of Figures~\ref{fig:gvminModel1} \& \ref{fig:gvminModel2}, we show the difference in $g(v_{\rm min})$ at the two dates at which the isolated MW has its extrema.  The relative speed between the DM and the Earth is highest in June and lowest in December, leading to an excess of high-speed particles ($v \gtrsim 200$ km/s) in June and low-speed particles in December (see also Figure~\ref{fig:Earth}).  The difference in $g(v_{\rm min})$ is proportional to the amplitude of the annual modulation of the direct-detection recoil spectrum. 

Because the LMC leads to a boost in highest-speed particles (both of LMC origin and MW particles accelerated by the passage of the LMC), the post-LMC-impact annual modulation of $g(v_{\rm min})$ largely follows that of the isolated MW below speeds of 600 km/s for both the isotropic (Model 1) and anisotropic (Model 2) MW cases. There is a slight shift in the zero point of the annual modulation amplitude, as indicated by the sharp excursions around 200 km/s in the bottom right panels of Figures~\ref{fig:gvminModel1} \& \ref{fig:gvminModel2}.  Above 600 km/s, the annual modulation amplitude rises significantly relative to the isolated MW case.  This pattern of annual modulation amplitude as a function of $v_{\rm min}$ is unique to the LMC, as other dark-matter substructures have different $g(v_{\rm min})$ shapes and annual modulation amplitudes \cite{Freese:2013}.

\begin{figure}[tbp]
\centering
\mbox{\includegraphics[width=3in, trim={0in 0.5in 0.7in 0.6in}, clip]{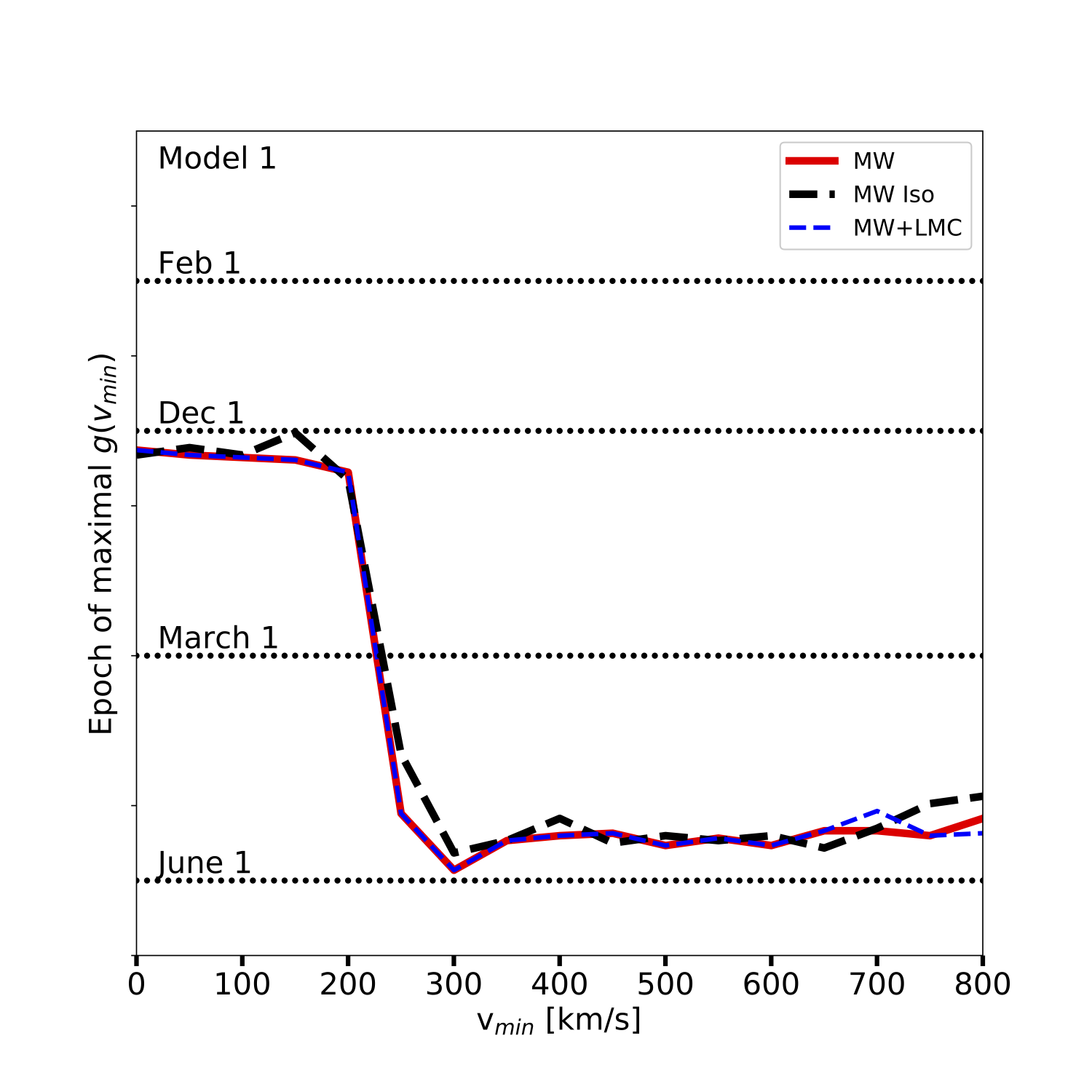}
\includegraphics[width=3in, trim={0in 0.5in 0.7in 0.6in}, clip]{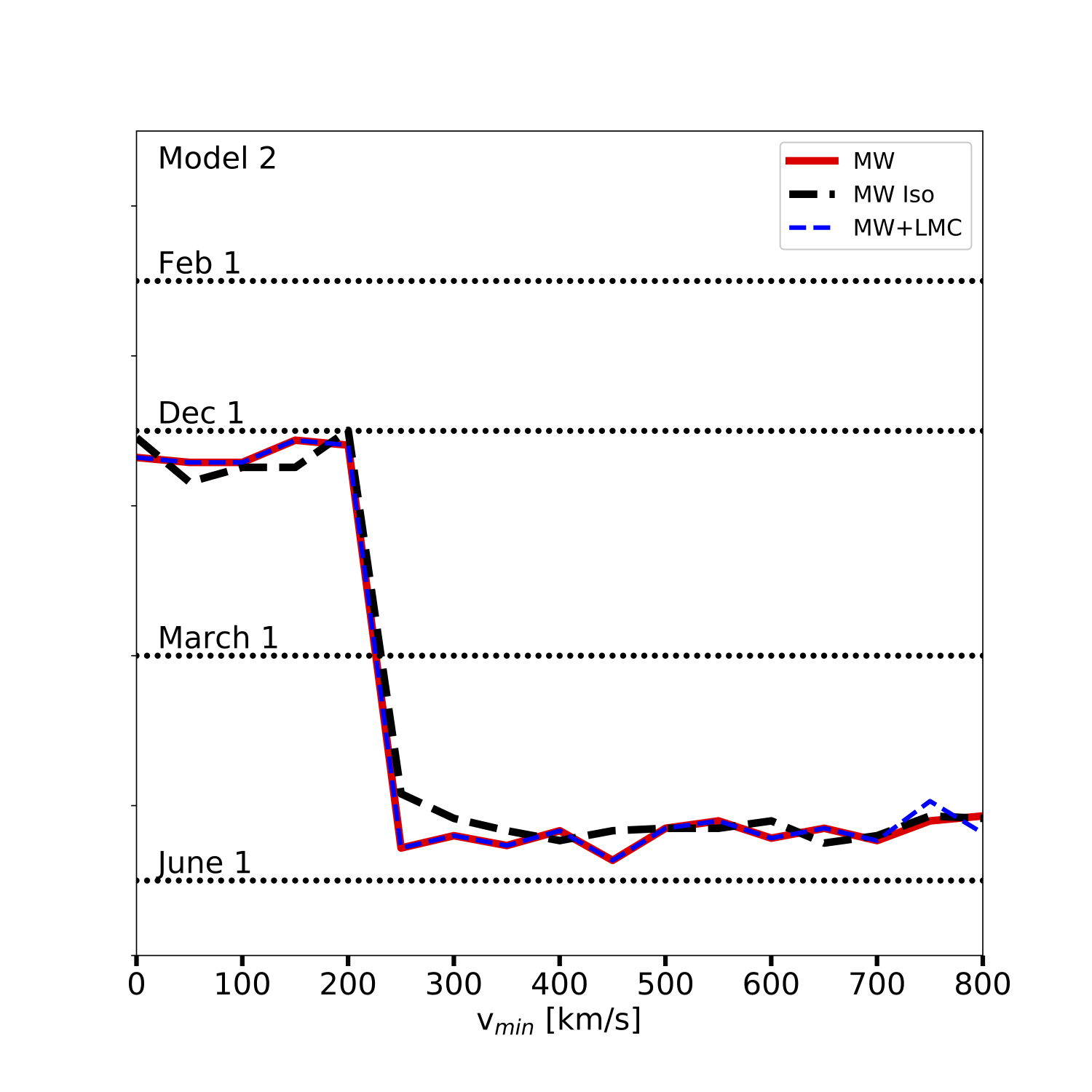}}
\caption{\label{fig:gvminPhase}Phase of $g(v_{\rm min})$: plotted is the date where $g(v_{\rm min})$
is maximized, accounting 
for the annular motion of the Earth. Results are plotted for Model 1 (left) and Model 2 (right). At high speeds the event rate is expected to be maximized in June.} 
\end{figure}

Because the LMC's velocity vector is almost exactly anti-parallel to the Sun's motion, the phase of annual modulation largely mimics that of the isolated MW, unlike other possible DM kinematic substructures.  This point is illustrated in Figure~\ref{fig:gvminPhase}, we show the date at which $g(v_{\rm min})$ is maximized as a function of $v_{\rm min}$.  We have included the effect of the relative speed of the Earth with respect to the local WIMP population, but not gravitational focusing, which shifts the phase for speeds $v_{\rm min} \lesssim 300$ km/s \cite{Lee2014}, below the speeds of LMC and LMC-induced MW accelerated particle populations.  Thus, the Figure shows that the signature of the LMC in annual modulation measurements lies not in its phase but in its amplitude.

\begin{figure}[tbp]
\centering
\mbox{\includegraphics[width=4in, trim={0in 0in 0.5in 1in}, clip]{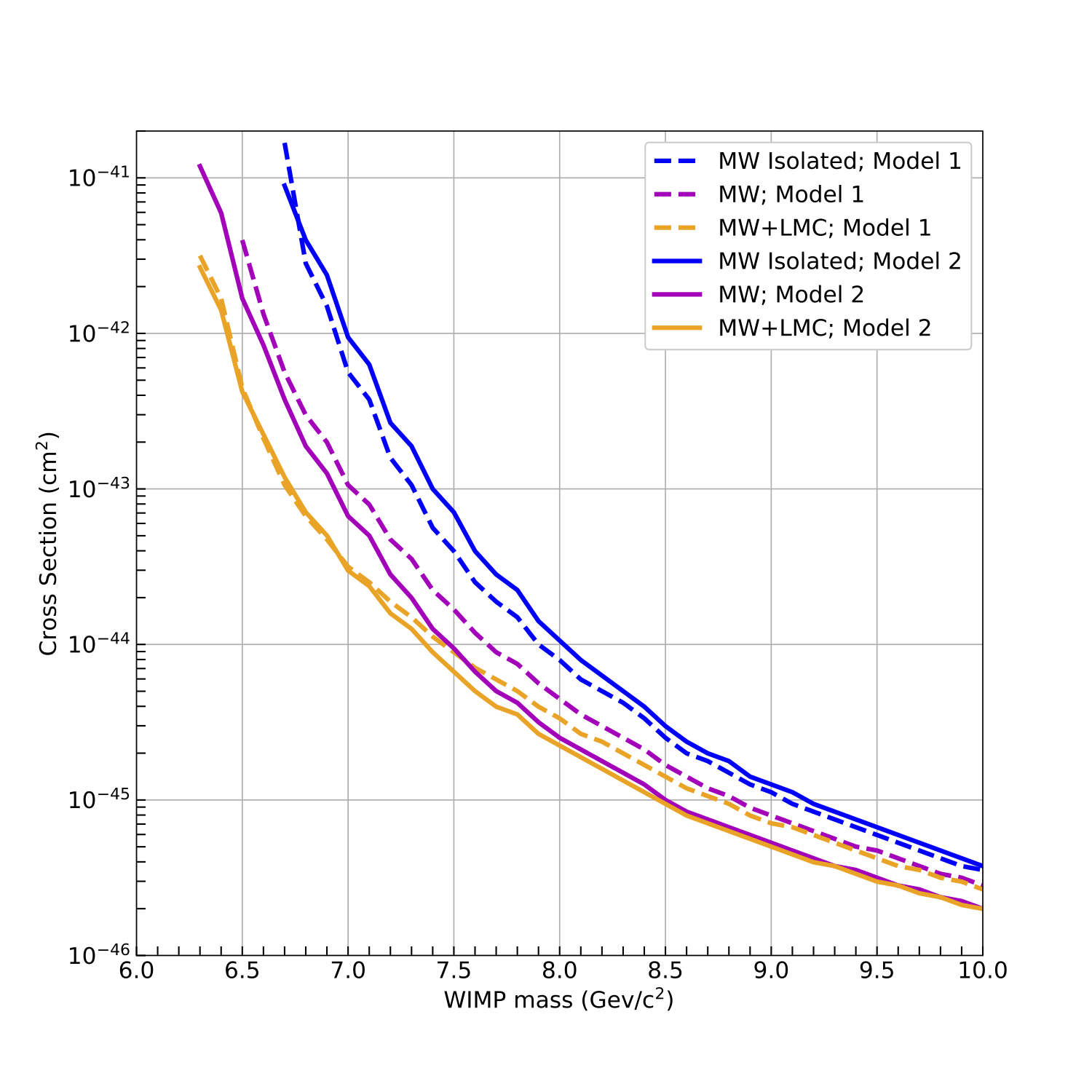}}
\caption{Limits on the spin-independent DM-nucleon scattering cross section \sigmapsi, assuming a Xenon target with an exposure of 1 kton per year and a 4.9 keV energy threshold. The plot focuses on the lowest mass WIMPS. The inclusion of the LMC results in cross section limits that are significantly lower at low DM particle mass than found in the SHM (roughly equivalent to the results for the isolated MW: MW Isolated, blue lines).  This figure illustrates that the cross section limits are affected by both accelerated MW DM halo particles (MW: magenta lines) and also by LMC particles (LMC+MW: Orange). Cross section limits are lower in Model 2 (radially anistropic DM halo) than in Model 1 (isotropic), but are comparable at the lowest DM masses.   \label{fig:CrossSection} }
\end{figure}

The key effect of the LMC on direct-detection experiments is to extend the reach of those experiments to lower WIMP mass.  The existence of a low-energy threshold implies that there is a minimum WIMP mass to which experiments are sensitive, given a fixed speed distribution.  Figure~\ref{fig:CrossSection} illustrates the resulting limits on the spin-independent
DM-nucleon scattering cross section \sigmapsi~at low DM particle mass compared to that assuming an isolated MW.  In this figure, the limits correspond to the results from the Xenon-1T 1 ton-year exposure result \cite{Aprile:2018}.  We translate the experimental limit from that work to more general forms of the WIMP speed distribution function according to the method of Ref. \cite{Necib:2019iwb}.  The presence of the LMC causes the cross-section limits to shift downward and to lower WIMP masses. Importantly, the effect is independent of the assumed MW dark matter velocity anisotropy at the lowest masses. 

\begin{figure}[tbp]
\centering
\mbox{\includegraphics[width=6in, trim={0in 0.3in 0in 0in}, clip]{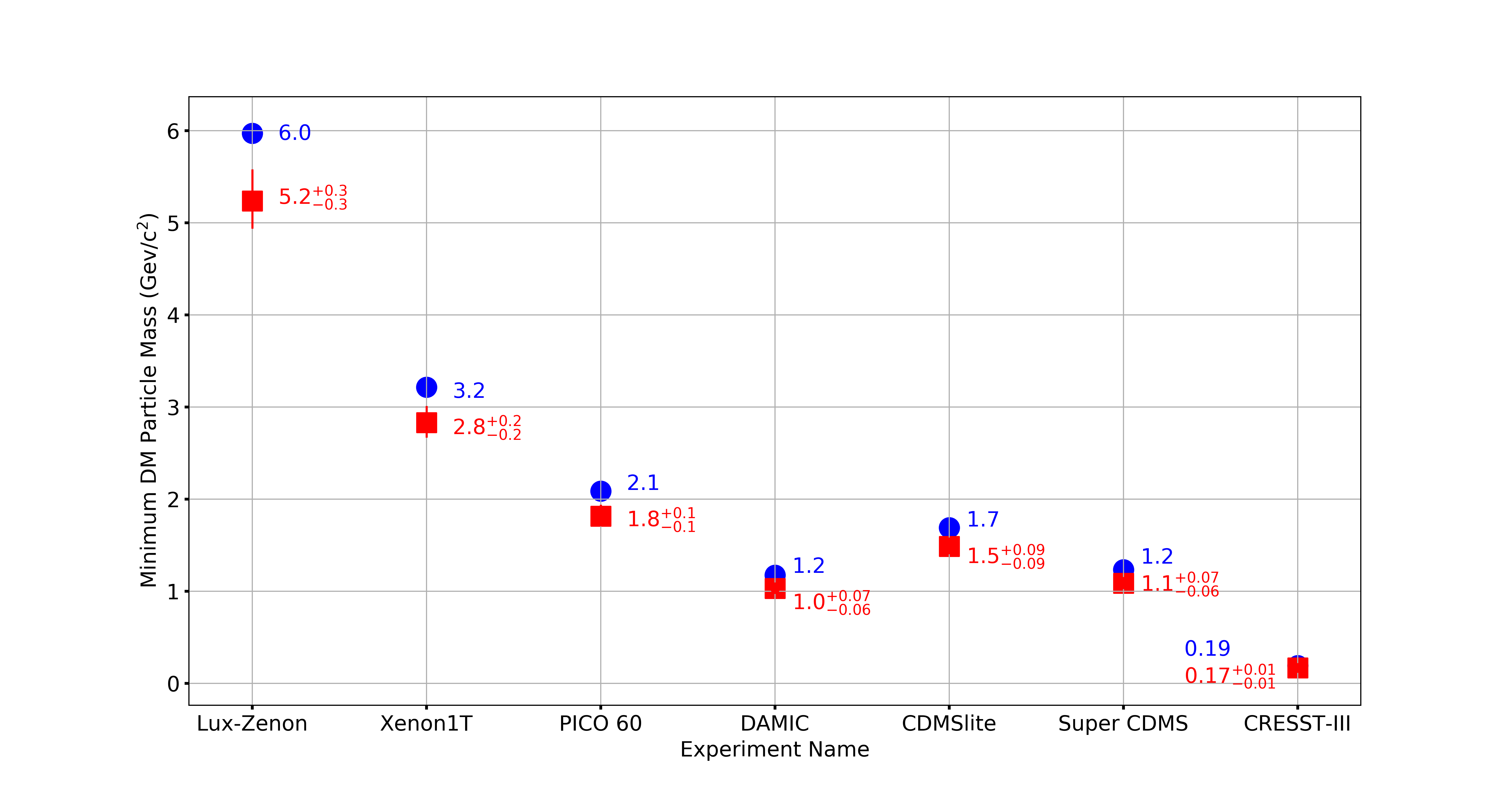}}
\caption{Minimum DM particle mass reachable by the listed experiments, ordered by descending 
energy threshold ($Q_{min}$), when the LMC is included (red squares, $v_{\rm min}$ =850$^{900}_{800}$ km/s) or ignored (blue circles, $v_{\rm min}$ =750 km/s). Thresholds
and target nuclei are described in the caption of Figure~\ref{fig:gvminModel1}, with the addition of CDMSlite ($Q_{min}$=0.5 keV, Germanium target) \citep{Agnese:2018}. The LMC extends the reach of experiments to lower WIMP mass. \label{fig:DMmin} }
\end{figure}

Because of the high-speed WIMP population induced by the LMC, the cross section limits can be significantly lower for low-mass WIMPs (by over an order of magnitude) than expected 
when the LMC is neglected. Furthermore, some experiments are more sensitive to low-mass WIMPs than they currently claim to be.  This is illustrated for specific experiments in Figure~\ref{fig:DMmin}, where we compare the WIMP mass that may be probed with DM at the MW escape speed relative to the lower WIMP mass that is accessible because of the LMC-induced high-speed tail in the Solar Neighborhood.

\section{Discussion}
\label{sec:Discussion}

In this work, we have shown that the LMC is responsible for the high-speed tail of the WIMP velocity distribution in the Solar Neighborhood, and that this extends the reach of direct-detection experiments to lower WIMP mass and lower cross section.  In this section, we discuss our findings in the context of the uncertainty in the LMC's and MW's total DM masses, and in recent work in reconstructing the Solar Neighborhood WIMP phase-space density using the stellar halo.

\subsection{Impact of Uncertainties in the LMC and MW Mass and Orbit}
\label{sec:LMCMass}

In this study we have utilized a massive LMC halo model with infall mass of $2.5\times10^{11}$ M$_\odot$. Here, we explore the impact of the uncertainty on the LMC's and MW's halo masses on our results.

We include three different LMC infall halo mass models, as introduced in G19:  $10^{11}$ M$_\odot$ (LMC3), $1.8\times 10^{11}$ M$_\odot$ (LMC2), $2.5\times 10^{11}$ M$_\odot$ (LMC1; Fiducial). These LMC mass models encompass a representative range for the expected mass of the LMC {\it prior} to entry within the virial radius of the MW $\sim$2 Gyr ago (see discussion in G19 and \S\ref{sec:massivesat}).  Note that all results presented in prior sections have focused on a fiducial (LMC1) mass model. 

\begin{figure}[tbp]
\centering
\mbox{\includegraphics[width=4in, trim={0in 0in 0.5in 0.5in}, clip]{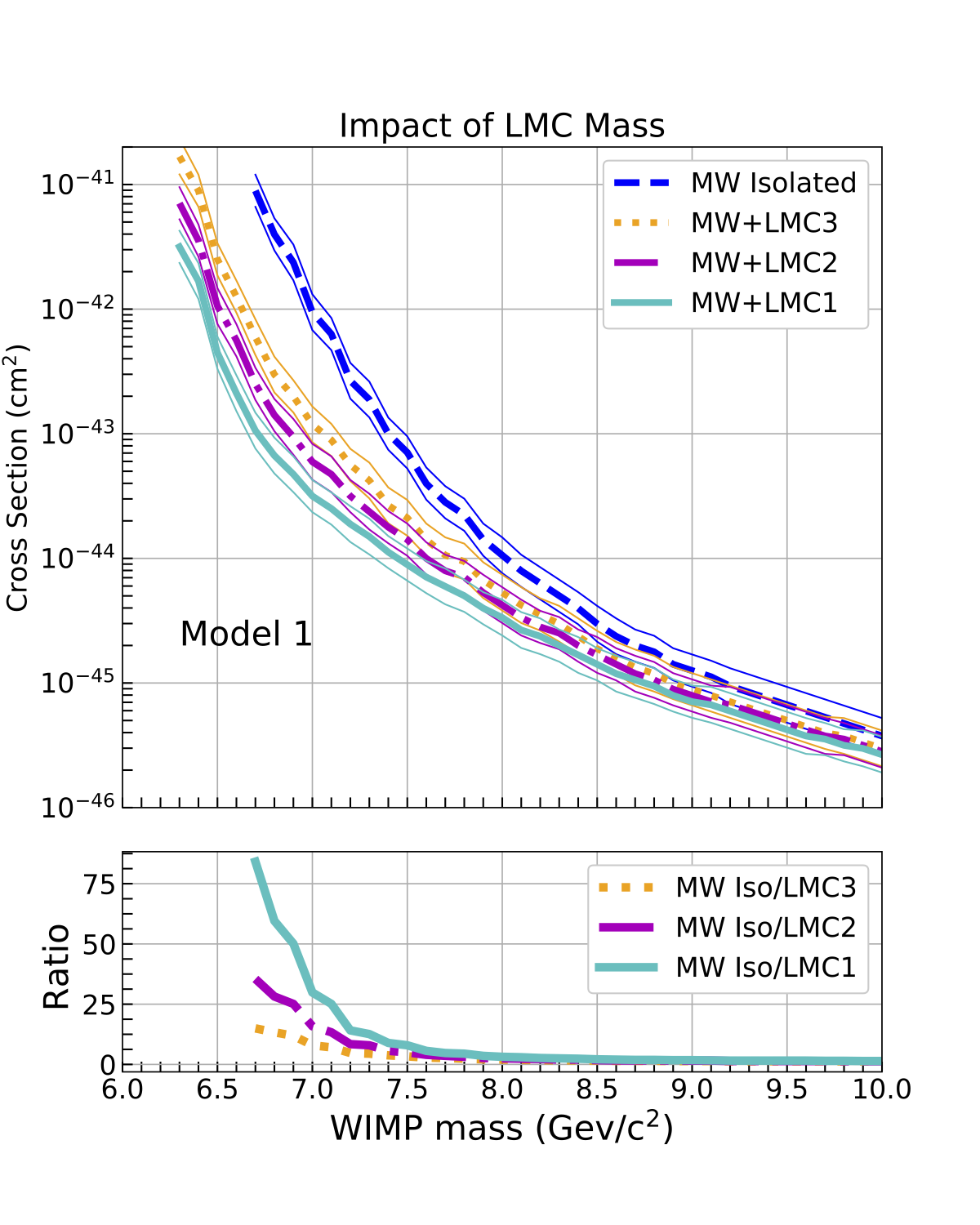}}
\caption{Impact of LMC Mass on the cross section. LMC mass ranges explored: $10^{11}$ M$_\odot$ (LMC3, dotted), $1.8\times 10^{11}$ M$_\odot$ (LMC2, dashed-dotted), $2.5\times 10^{11}$ M$_\odot$ (LMC1; Fiducial, solid). The MW mass model is Model 1 in each case; Model 2 results in similar trends. The cross section limits come from the same analysis of Xenon-1T limits used in Figure~\ref{fig:CrossSection}. The bottom panel shows the ratio between the cross section of the MW in isolation (no LMC) relative to the result of the MW model including each LMC mass model, as marked. Even in the lowest LMC mass model (LMC3) the cross section is still impacted by an order of magnitude. \label{fig:DiffCrossSection} }
\end{figure}

The top panel of Figure~\ref{fig:DiffCrossSection} illustrates the resulting cross section for each LMC model using MW Model 1. The bottom panel illustrates the ratio of these cross sections to the MW in isolation.  The cross section is impacted at low DM masses in all cases -- general trends are the same provided the LMC mass at infall is at least $10^{11}$ M$_\odot$.  The change in cross section owes to the increased
 number of LMC particles in the 
Solar Neighborhood as a function of increasing LMC infall mass.  The normalized speed distribution of LMC particles does not change significantly with increasing LMC mass, rather, there are simply more particles at all speeds. 
For example, LMC1 has roughly a factor of 10 more particles at speeds in excess of 850 km/s than LMC3. 

Note that adopting a different density profile for the LMC's DM halo at infall can also change the amount of LMC 
particles in the Solar Neighborhood. The NFW density profile, for instance, falls off less rapidly than the Hernquist profile with distance, which would increase the amount of LMC particles. This may allow for lower LMC infall masses than studied here. Our results assume that, at a minimum, the LMC's halo extends for 50 kpc (which is true for all three LMC models explored in this study). This enables the LMC's halo to overlap with the Solar Neighborhood at the present day, regardless of the 
slope of the profile. 

In contrast, the number of accelerated MW particles does not change significantly with increasing LMC mass.
This is consistent with the behavior of the halo response to the LMC's orbit, which does not change in amplitude 
as a function of LMC mass at Galactocentric distances $<$ 70 kpc (see right panel of G19 Figure 25). 
The MW particles accelerated by the LMC are an important contributor to the high speed tail - their properties appear to be independent of the exact LMC profile.

The direction of motion of LMC particles in the Solar Neighborhood is dependent on the velocity vector of the LMC at closest approach to the MW ($\sim$48 kpc), which is model-independent. In other words, changing either the mass or density profile of the LMC or MW's halo does little to change the distance, velocity vector or timing of the LMC's closest approach \citep{Kallivayalil:2013, Salem:2015}. Calculations presented here about the impact of LMC particles to the Solar Neighborhood are thus robust to 
uncertainties in the orbit of the LMC, as these are negligible over the past 1 Gyr \citep{Patel:2017}.

The total DM mass and density profile of the MW may impact the amplitude of the halo response to the passage of the LMC. Luckily, both are well-constrained by the rotation curve within 20 kpc. As demonstrated by the exercise of changing the LMC mass, changes in the amplitude of the halo response at large scales, $>$50 kpc, do little to impact the halo response in the inner halo (G19). More significant are uncertainties in the velocity anisotropy profile, which change the {\it initial} velocity distribution in the Solar Neighborhood before the LMC arrives. We have studied this effect in our analysis of Model 1 and Model 2, finding that a radially biased velocity distribution can promote the acceleration of MW particles. Note that 
changing the nature of the DM particle (e.g. ``fuzzy DM'' vs. cold dark matter)
may change the MW's halo response (see discussion in G19), and is the 
subject of future work.

\subsection{DM contribution of known substructure and debris flow vs. the LMC to the Solar Neighborhood} 

In addition to the virialized component of the MW halo (stellar and DM), several other un-phase-mixed structures certainly contribute to the distribution of stars in the Solar Neighborhood, and perhaps also to the DM.  These non-equilibrium structures -- gravitationally bound satellite galaxies and subhalos, narrow streams of dissolved satellites, and the spatially coherent but kinematically unmixed ``debris flow'' particles from dissolved or dissolving satellites \cite{Lisanti:2011as} -- may all lead to unique kinematic signatures in direct-detection experiments.  In this section, we place the LMC's implications for direct-detection experiments in this context, and how various authors infer properties of the Solar Neighborhood DM phase-space density based on stellar signatures of disequilibrium in the MW halo. We note that in our work, in addition to the LMC's DM debris, there is also a contribution from MW DM halo particles that have been accelerated through a Collective Response of the halo 
to the passage of the LMC \citep{G19}. These accelerated particles are distinct from the traditional definition of 
debris flow.

The most famous non-equilibrium structure is the Sagittarius dwarf spheroidal galaxy (Sag dSph).  This galaxy is moving in an orbital plane orthogonal to the 
MW's disk, perpendicular to that of the LMC.  It has two massive stellar streams wrapped completely around the MW \cite{Newberg:2001sx,Ibata2001,Majewski2003}. 
The associated stellar and DM debris stream is expected to exhibit speeds of 
400-500 km/s \cite{Purcell:2012sh} with respect to the Earth (see also Ref. \cite{Freese:2013,Freese:2003tt}, who find stream debris to contribute at $\sim$350 km/s). These authors find that
the SHM is insufficient to describe the non-Maxwellian velocity distribution caused by Sag dSph
and that the presence of Sag dSph may reduce the amplitude of the annual modulation of the event rate by as much as factor of two.
Furthermore, Ref. \cite{Purcell:2012sh} find that although Sag dSph may contribute more to the Solar Neighborhood DM density than the LMC does ($1-2\%$ vs $\sim 0.2\%$), the debris enhances event rates by only $\sim 20\%$ because the speed distribution of Sag dSph particles lies close to the peak of that of the overwhelmingly dominant MW population.   Thus, even though the LMC contributes less to the Solar Neighborhood DM density than Sag dSph does, the high-speed nature of LMC and LMC-accelerated MW particles leads to a much stronger enhancement of events in direct-detection experiments relative to isolated MW predictions in the relevant energy window.

Recently, multiple authors have identified additional stellar substructures 
within the inner halo of the MW. These data are
consistent with a scenario in which our Galaxy has undergone one or two  
substantial merger events $\sim$9-10 Gyr ago:
1) Gaia-Sausage, characterized by radially anisotropic stellar debris \citep{Belokurov2018} ; 2) retrograde 
stars, such as the S1 stream and remnants of the Sequoia dwarf galaxy \citep{Myeong2019}; 3) Gaia-Enceladus debris that likely consists of both Gaia-Sausage stars and the retrograde components of Sequoia \citep{Helmi2018,Kruijssen2019}; and 4) the prograde stream Nyx \cite{Necib2019}. These features lie somewhere in the continuum of distinct streams and debris flow. 

These substructures suggest
that the local DM halo
is not in equilibrium and will deviate from SHM \cite{Necib:2019iwb}. 
In particular, the kinematics and distribution of Gaia-Sausage stars are best matched by the disruption of a system on a radial orbit, resulting in a strongly radially anisotropic velocity distribution of stars in the solar neighborhood, which may have implications for directional direct-detection experiments \citep{Evans:2019bqy}. The Gaia-Sausage and other debris-flow material may contribute at the $\sim 10\%$ level to the Solar Neighborhood DM density, but the velocity distribution of stars associated with the Gaia-Sausage populate speeds of $v_{\rm min} <$500 km/s \citep{Belokurov2018,Evans:2019bqy,Necib:2018igl}, much lower than the expected LMC debris flow.

The highest energy stars in the halo are typically retrograde \citep{Myeong2018,Myeong2018b,Myeong2018c,Helmi:2017}, suggesting these stars were accreted through past mergers \citep{Quinn:1986,Norris:1989,Carollo:2007, Beers:2012}.  The S1 Stream is also a retrograde structure and Ref. \citep{Myeong2019}
suggest a common link with a single dwarf galaxy progenitor, Sequoia, which was potentially 
accreted $\sim$9 Gyr ago, bringing $\sim$10$^{10}$ M$_\odot$ worth of DM.  
The retrograde S1 stream is estimated to be significant at speeds of 450-650 km/s with respect
to the Earth \citep{Buckley:2019, OHare:2018trr}, causing an excess for recoil energies at $v_{\rm min} \sim 550$ km/s. 

While the kinematics associated with all of the listed substructures are expected to cause deviations in the local debris flow from a Maxwellian distribution, none of these structures impact the velocity distribution at Earth-frame-speeds as high as those expected for the LMC debris flow ($>$700 km/s). In fact, the DM distribution inferred from SDSS-Gaia DR2 suggests that there are far fewer high-speed particles than expected from the SHM \citep{Necib:2019iwb}.
{\it The LMC is thus expected to be the most significant contributor to the high-speed local DM debris flow.}

The inferences about Solar Neighborhood DM substructure based on the kinematic substructure in stellar halo stars depends on the assumption that stars trace the kinematic structure of DM.  Indeed, Ref. \cite{Evans:2019bqy} and others argue that the Solar Neighborhood DM can be modeled with the SHM with the addition of DM associated with stellar substructures.  Ref. \cite{Necib:2018igl} suggest that the strongest correlation between stars and DM exists for the relaxed component of the halo and the debris flow. However, relating the stellar and DM components of unmixed and dissolving substructures is less straightforward \cite{Bozorgnia2019}. DM halos are vastly more extended than the stellar component of satellite galaxies, and typically at least 90\% of dark matter must be tidally stripped before the stars are touched \cite{Penarrubia2008}.  Thus, in detail, some of the inferences of DM kinematic substructure in the Solar Neighborhood may change as we understand the relative kinematics of stars and DM in disrupting massive satellites \citep{Necib:2019iwb,Bozorgnia2019}. 

Significantly, relying on stars to trace DM in the Solar Neighborhood misses any DM component arising from satellites whose stellar component is intact, like the LMC's.  The LMC contribution to the local DM debris flow will not have a stellar counterpart -- 
LMC DM particles are from the outskirts of its halo ($\sim$ 50 kpc from the center of the LMC). 
Furthermore, no simulated MW stellar 
disk particles in the Solar Neighborhood reach geocentric speeds in excess of 750 km/s at the present day.
This is because the kinematics of MW stellar disk particles in the Solar Neighborhood are not affected by the resonances that impact the DM halo, such as the Collective Response (G19). Note, however, that the outer stellar disk of the MW ($>$15 kpc) is impacted by the LMC, inducing a pronounced warp \citep{Laporte:2018}. Similarly, we do not expect the LMC to dramatically alter the local kinematics of a ``dark disk'' of DM that is kinematically similar to the stellar disk (whether it be of accretion origin \cite{Read:2008fh} or a result of strong dissipation in the hidden sector \cite{Fan2013}).

Thus, the kinematic structure in the Solar Neighborhood arising from the LMC is unique -- it cannot be traced by stars, and its super-escape-velocity average speed is much higher than any other known component of the Solar Neighborhood.

\section{Conclusions}

A positive detection of DM interactions in a laboratory is the holy grail of particle physicists.  Of particular interest for the WIMP/CDM paradigm is a detection of Galactic DM as it passes through underground laboratories.  The next generation of multi-ton direct-detection experiments will test conventional WIMP-nucleon interaction cross sections down to the irreducible background of solar and atmospheric neutrinos \cite{Cushman:2013}.

In this study, we utilize a suite of controlled simulations of the recent infall of a massive ($1-2.5 \times 10^{11}$ M$_\odot$) LMC into the MW's DM halo to study the resulting velocity distribution of DM particles in the Solar Neighborhood \cite{G19}. These simulations account for three effects simultaneously: 1) the MW's disk potential; 2) the DM halo velocity anisotropy profile; and, for the first time, 3) the presence of a massive LMC on first infall \cite{Besla:2007,Besla:2010}.

We find that the high-speed tail of the local DM distribution is 
overwhelmingly of LMC origin -- both as particles that were once bound to the LMC and 
MW halo particles that have been accelerated owing to halo resonances induced by 
the motion of the LMC \cite{G19}. 
The general trends we report below require that
the LMC mass at infall is at least $10^{11}$ M$_\odot$, as expected from CDM cosmology. 
Although, we note that changing the LMC's DM halo profile from the Hernquist profile adopted here 
to the more extended NFW profile may allow for more LMC particles in the Solar Neighborhood even in lower mass LMC models.

\begin{itemize}
\item The LMC both directly contributes particles and also 
accelerates local MW DM halo particles to high speeds, reaching velocities 
$>700$ km/s, with respect to the Earth.  These are significantly higher speeds 
than traditionally assumed in the SHM.  

\item The velocity vector of these high-speed particles reflect 
 the direction of motion of the LMC at closest approach, which is roughly coincident with the reflex motion of the Sun. The spatial distribution of high speed particle trajectories
 is distinct from that of the SHM model or other substructures, such as those seen in Gaia \citep{Evans:2019bqy}. It is more concentrated and skewed towards the South.

\item The LMC is found to increase the 
the time-averaged inverse speed, $g(v_{\rm min})$, resulting in 
an expected increase in the event rate at speeds $>700$ km/s by factors of a few or more.  

\item The LMC causes the limits on the spin-independent
DM-nucleon scattering cross section, \sigmapsi,
to be significantly lower for low mass WIMPS than expected 
when the LMC is neglected (e.g. SHM Model). The effect will be even more dramatic for inelastic dark matter models \cite{smith2001,Kuhlen:2010}, or for models of elastic scattering with other types of potentials \cite{fan2010,fitzpatrick2013}.

\item The LMC accelerates MW DM particles to higher speeds if the initial
velocity anisotropy profile of the MW is radially biased (Model 2 vs. Model 1).
This is in agreement with the augmentation of the MW's halo response to the LMC's orbit seen 
in G19 using cosmologically motivated, radially biased halo models.

\end{itemize}

The LMC extends the high velocity tail of the local DM distribution to higher speeds than previously considered in SHM or un-phase-mixed models (Sagittarius or Gaia substructures) owing to the fact that the LMC is on first infall (high energy orbit) with a trajectory that is 
primarily in the same direction as the reflex motion of the Sun. 
Significantly, the LMC contribution to the local DM debris flow will not have a stellar
counterpart. 
The LMC thus 
uniquely impacts the prospects of detecting low mass WIMPS ($<$10 GeV/c$^2$) in current and future 
direct-detection experiments.

\acknowledgments

We would like to thank Nassim Bozorgnia, Mariangela Lisanti, 
Lina Necib, David Spergel and Louis Strigari for helpful comments 
that have improved this manuscript.
This work is supported by a NASA ATP grant 17-ATP17-0006 to GB, and by NSF Grant No. AST-1615838 to AHGP.

\bibliographystyle{JHEP}
\bibliography{main}{}

\end{document}